\documentclass[sigconf]{acmart}

\usepackage{graphicx}
\usepackage{hyperref}
\usepackage{multirow}
\usepackage{booktabs}
\usepackage{verbatim}
\usepackage{longtable}
\usepackage{enumitem}
\usepackage{array}
\usepackage{changepage}
\usepackage{listings}
\usepackage{float}

\usepackage{xcolor}

\definecolor{RQ1}{HTML}{4061b3}
\definecolor{RQ2}{HTML}{b85455}
\definecolor{RQ3}{HTML}{358f7f}
\definecolor{RQ1Background}{HTML}{f1f7ff}
\definecolor{RQ2Background}{HTML}{fbf7f2}
\definecolor{RQ3Background}{HTML}{f1fff2}

\newcommand{\reflexivecolor}[1]{\textcolor{RQ1}{#1}}
\newcommand{\transparentcolor}[1]{\textcolor{RQ2}{#1}}
\newcommand{\collaborationcolor}[1]{\textcolor{RQ3}{#1}}

\newcommand{\code}[1]{%
  \colorbox{black!10}{\texttt{#1}}%
}

\newcommand{\reflexivefeature}[1]{%
  \colorbox{RQ1Background}{\texttt{#1}}%
}

\newcommand{\transparentfeature}[1]{%
  \colorbox{RQ2Background}{\texttt{#1}}%
}

\newcommand{\collaborationfeature}[1]{%
  \colorbox{RQ3Background}{\texttt{#1}}%
}

\AtBeginDocument{%
  }

\copyrightyear{2026}
\acmYear{2026}
\setcopyright{cc}
\setcctype{by-nc-nd}
\acmConference[CHI '26]{Proceedings of the 2026 CHI Conference on Human Factors in Computing Systems}{April 13--17, 2026}{Barcelona, Spain}
\acmBooktitle{Proceedings of the 2026 CHI Conference on Human Factors in Computing Systems (CHI '26), April 13--17, 2026, Barcelona, Spain}
\acmPrice{}
\acmDOI{10.1145/3772318.3791275}
\acmISBN{979-8-4007-2278-3/2026/04}

\newcommand{\sysname}{\textsc{Reflexis}}

\begin{document}

\begin{teaserfigure}
  \centering
  \includegraphics[width=.8\textwidth]{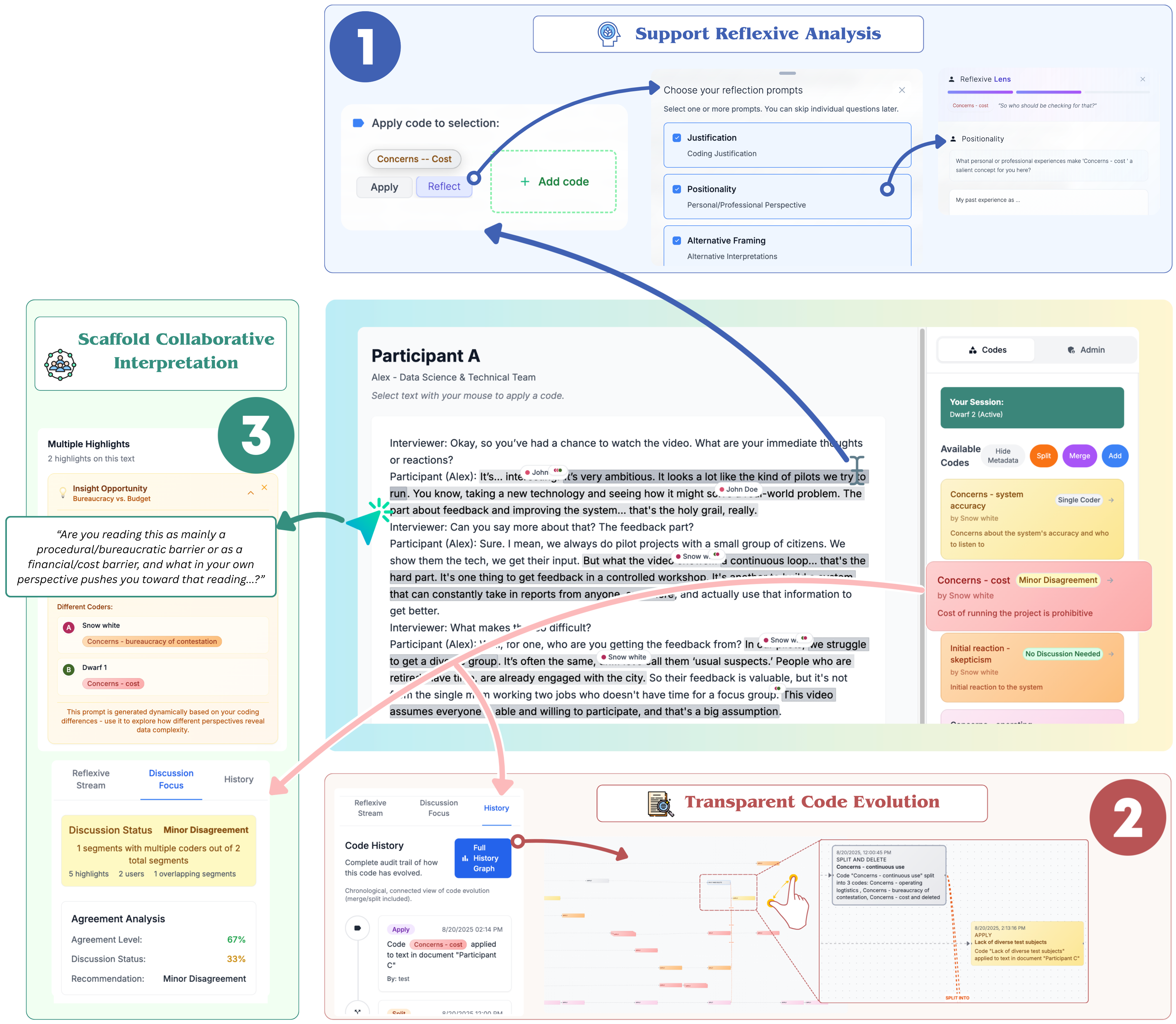}
  \caption{
  \sysname{}: a collaborative reflexive thematic analysis workspace. \sysname{} supports reflexive collaborative thematic analysis by weaving researcher reflexivity, methodological transparency, and collaborative interpretation into a single, integrated workflow. To support \reflexivecolor{\textbf{reflexive analysis (1)}}, it grounds the researcher in reflexive practice by prompting them to articulate their positionality and offering tools for in-situ reflection during coding. To ensure this process is transparent and tangible, the system provides a \transparentcolor{\textbf{transparent code evolution (2)}} via detailed analysis histories and automated code drift detection, which promotes more robust and insightful analysis. Finally, the system \collaborationcolor{\textbf{scaffolds collaborative interpretation (3)}}, with positionality-aware prompts and discussion foci that transform interpretive differences during collaboration into productive dialogue. 
  }
  \Description{TODO}
  \label{fig:teaser}
\end{teaserfigure}

\title{\sysname{}: Supporting Reflexivity and Rigor in Collaborative Qualitative Analysis through Design for Deliberation}


\author{Runlong Ye}
\affiliation{
  \institution{Computer Science, \\University of Toronto}
  \city{Toronto}
  \state{Ontario}
  \country{Canada}
}
\orcid{0000-0003-1064-2333}
\email{harryye@cs.toronto.edu}

\author{Oliver Huang}
\affiliation{
  \institution{Computer Science, \\University of Toronto}
  \city{Toronto}
  \state{Ontario}
  \country{Canada}
}
\orcid{0009-0007-1585-1229}
\email{oliver@cs.toronto.edu}

\author{Patrick Yung Kang Lee}
\affiliation{
  \institution{Computer Science, \\University of Toronto}
  \city{Toronto}
  \state{Ontario}
  \country{Canada}
}
\orcid{0000-0002-3385-5756}
\email{patricklee@cs.toronto.edu}

\author{Michael Liut}
\affiliation{
  \institution{Mathematical and Computational Sciences, \\University of Toronto Mississauga}
  \city{Mississauga}
  \state{Ontario}
  \country{Canada}
}
\orcid{0000-0003-2965-5302}
\email{michael.liut@utoronto.ca}

\author{Carolina Nobre}
\affiliation{
  \institution{Computer Science, \\University of Toronto}
  \city{Toronto}
  \state{Ontario}
  \country{Canada}
}
\orcid{0000-0002-2892-0509}
\email{cnobre@cs.toronto.edu}

\author{Ha-Kyung Kong}
\affiliation{
    \institution{School of Information, \\Rochester Institute of Technology}
    \city{Rochester} 
    \state{New York}
    \country{USA}
}
\orcid{0000-0001-8974-8610}
\email{hidy.kong@rit.edu}


\begin{abstract}
Reflexive Thematic Analysis (RTA) is a critical method for generating deep interpretive insights. Yet its core tenets, including researcher reflexivity, tangible analytical evolution, and productive disagreement, are often poorly supported by software tools that prioritize speed and consensus over interpretive depth. To address this gap, we introduce \sysname{}, a collaborative workspace that centers these practices. It supports reflexivity by integrating in-situ reflection prompts, makes code evolution transparent and tangible, and scaffolds collaborative interpretation by turning differences into productive, positionality-aware dialogue. Results from our paired-analyst study ($N=12$) indicate that \sysname{} encouraged participants toward more granular reflection and reframed disagreements as productive conversations. The evaluation also surfaced key design tensions, including a desire for higher-level, networked memos and more user control over the timing of proactive alerts. \sysname{} contributes a design framework for tools that prioritize rigor and transparency to support deep, collaborative interpretation in an age of automation.
\end{abstract}




\maketitle

\section{Introduction}

Qualitative research is foundational across disciplines, including health, sociology, and education, for generating in-depth interpretations of complex human experiences \cite{willig2017sage, braun2019reflecting}. Building on this broader tradition, \textit{Reflexive Thematic Analysis} (RTA) has become a widely adopted approach within HCI and CSCW because it treats researcher subjectivity not as a bias to be minimized, but as a central analytic resource. Unlike methods focused on ``coding reliability'' and objective measurement, RTA prioritizes deep, theory-building practice grounded in the researcher’s situated judgments \cite{braun2021one, mcdonald2019reliability}. This interpretive stance is critical for producing impactful findings, acknowledging that human experience is a complex reality to be understood rather than a simple phenomenon to be measured.

While RTA theoretically centers the researcher's positionality and the evolution of meaning \cite{probst2014double}, achieving this rigor in practice remains difficult. In current workflows, self-reflexive notes are often disconnected from the data; the evolving nature of codes is rarely documented with provenance; and collaborator positionalities remain invisible when interpretations diverge \cite{singh2025exploring}. 

Compounding this challenge is the rapid growth of tool support for qualitative data analysis (QDA) that has largely pulled in the opposite direction. While commercial tools like NVivo and ATLAS.ti facilitate management and prioritize quantitative insights, recent research systems leveraging Large Language Models (LLMs) have prioritized \textit{automation, scalability, and convergence}. New tools often act as ``auto-coders'' to reduce human effort \cite{CoAIcoder, scholarmate} or mediate discussions to reach rapid consensus \cite{CollabCoder}. 

These prevailing designs implicitly optimize for speed and agreement, often inadvertently shrinking the interpretive space \cite{bowman2023using}. By smoothing over friction, they offer limited support for the practices that make RTA rigorous, such as sustained reflexive engagement, transparency regarding how interpretations change over time, and principled disagreement \cite{jiang_supporting_serendipity, schroeder_llms, soden2024evaluating}.

We address this misalignment with \sysname{}, a collaborative workspace designed to embed reflexivity, analytic history, and positionality directly into the analysis loop. Crucially, our goal is \textit{not} to automate the coding process for speed, \textit{nor} to enforce inter-coder reliability. Instead, we aim to scaffold methodological rigor by \emph{designing for deliberation} \cite{ma2025towards}. \sysname{} supports this through three principal mechanisms: \textbf{(1) In-situ Reflexivity:} Instead of separating memos from coding, \emph{in-situ reflective prompts} link researchers' evolving perspectives to specific data excerpts without disrupting their analytical flow. \textbf{(2) Transparent Provenance:} To counter the opacity of evolving concepts, a transparent \emph{analysis history} records the full provenance of every code's evolution (creations, splits, merges), while a \emph{Code Drift Alert} surfaces potential inconsistencies for human review. \textbf{(3) Principled Disagreement:} When interpretations diverge, \emph{positionality-aware collaboration} and \emph{Discussion Focus} surface these differences as opportunities for deeper theorization rather than errors to be fixed.

Consistent with RTA's interpretivist lens, all AI assistance in \sysname{} (e.g., detecting drift) is advisory and optional, ensuring the system supports, rather than supplanting, human interpretation. Guided by this stance, we investigate the effectiveness of \sysname{} in support of collaborative RTA via the following research questions:

\begin{adjustwidth}{2em}{}
\begin{enumerate}
    \item[\textbf{RQ1}] (Supporting Reflexive Analysis): How does integrating in-situ prompts support researchers in articulating their rationale and situating reflection within the coding workflow?

    \item[\textbf{RQ2}] (Making Code Evolution Transparent): How does visualizing code evolution and flagging drift influence researchers' perception of rigor and their immediate engagement with code definitions?
        
    \item[\textbf{RQ3}] (Scaffolding Collaborative Interpretation): How do positionality-aware scaffolds help research pairs identify and frame interpretive differences during synchronous collaborative work?
\end{enumerate}
\end{adjustwidth}

We conducted an evaluation study using \sysname{} as a conceptual probe with 12 qualitative researchers (6 pairs). Our findings demonstrate that \sysname{} transformed reflexivity from a summative activity into a granular, interconnected part of the coding workflow. Participants universally valued the system's ability to create a transparent, auditable history of code evolution. Furthermore, by automating the discovery and contextualization of interpretive differences, \sysname{} reframed disagreements not as conflicts to be resolved, but as structured, productive dialogues grounded in researcher positionality.

This paper contributes: \textbf{(1)} A conceptual framework for \emph{Design for Deliberation in Qualitative Analysis}, challenging the prevailing focus on automation and convergence by articulating how AI can be used to scaffold friction, reflexivity, and interpretive diversity. \textbf{(2)} \sysname{}, a collaborative system that operationalizes this approach within RTA. It introduces novel interaction mechanisms that transform implicit mental processes of reflection into explicit, tangible artifacts. \textbf{(3)} Empirical insights from an evaluation study demonstrating how deliberative designs shape analytical practice. Our work illustrates the value of \textit{positioning AI as a partner in rigor and deliberation}. Our findings suggest that this approach helps teams reframe disagreements as opportunities for deeper qualitative insights.
\section{Related Work}
\subsection{Reflexivity in Qualitative Analysis}

Qualitative methods are essential to HCI and CSCW for generating deep insights into the sociotechnical contexts of system use, which often fail to be captured by quantitative metrics \cite{mcdonald2019reliability}. Among these, thematic analysis (TA) has been increasingly adopted, due to its flexible approach to organizing and describing complex datasets \cite{bowman2023using}. A particularly influential form is reflexive thematic analysis (RTA), an interpretive method popularized by Braun and Clarke \cite{braun2006using, braun2013successful, braun2019reflecting}. RTA is a ``Big-Q'' qualitative method \cite{kidder1987qualitative}; it operates within an interpretivist paradigm that treats researcher subjectivity not as a bias to be minimized, but as a core part of the knowledge construction process, similar to approaches like constructivist grounded theory \cite{bowman2023using, charmaz2006constructing}. In RTA, the final analysis is understood to be at the intersection of the researcher and the data, shaped by the researcher’s positionality, which includes their background, values, and experiences, and grounded in their theoretical assumptions and coding practices \cite{braun2019reflecting, byrne2022worked, collins2018central, wicks2017coding}.

Despite its theoretical clarity, significant challenges arise in the application and evaluation of RTA within the field of human-computer interaction (HCI). A primary issue is the inconsistent reporting of reflexive practices. A recent review has found that most HCI papers claiming an RTA approach failed to adequately describe the researchers' positions, leaving readers to infer the interpretive lens through which the data was analyzed \cite{mcdonald2019reliability}. This lack of methodological transparency is a critical problem for qualitative research, where understanding the perspective of the analyst is paramount \cite{moravcsik2014transparency}. A related challenge is \emph{positivism creep}, where interpretivist work is judged by criteria from a positivist/quantitative research paradigm \cite{soden2024evaluating, bowman2023using}. Reviewers may request justifications for participant counts, demographic tables, or inter-coder reliability scores, metrics that are misaligned with interpretive validity \cite{braun2016mis, noble2015issues}. This pressure can result in ``performative'' positionality statements that do little to illuminate the research process \cite{singh2025exploring}. This environment places researchers in a difficult position, as they must advocate for the value of interpretation while navigating demands for positivist ``rigor'' \cite{watkins2017rapid}, a challenge that is intensified in collaborative team settings where managing reflexivity is inherently more complex \cite{barry1999using, probst2014double}.

\subsection{The Role of AI in Qualitative Analysis}

The HCI community has historically viewed qualitative analysis as resistant to computational solutions, given its labor-intensive nature and its reliance on interpretive rigor over positivist measurement \cite{jiang_supporting_serendipity}. Early visions for AI support were framed around principles of mixed-initiative interaction, where the system would act as a collaborator rather than a replacement for the researcher \cite{allen1999mixed, horvitz1999principles}. Initial attempts focused on semi-automating deductive analysis, where codes are derived from existing theory. These systems often struggled because the small datasets used in qualitative research provide insufficient training data \cite{yan2014optimizing, marathe2018semi}. With advances in NLP and the advent of LLMs, the focus has broadened to support inductive, ``bottom-up'' analysis, which is more characteristic of an interpretivist paradigm \cite{byrne2022worked}. Recent work has explored using LLMs for both deductive coding \cite{xiao2023supporting} and a wide array of other interpretive tasks, leading to a rapid expansion of new tools \cite{schroeder_llms, pang2025understanding}.

This development has populated the design space with a variety of systems for human-AI collaboration. Some tools focus on making large corpora more manageable by using topic modeling and clustering to help researchers find and curate relevant data subsets for manual analysis \cite{Scholastic, Teleoscope, NEEDLE, lam2024concept}. Others provide direct coding assistance by generating code suggestions with rationales \cite{PaTAT, SenseMate, rietz2021cody, wang2025lata}. More recent AI/LLM-powered systems emphasize the collaborative dimension of qualitative work, using AI to mediate discussion between human analysts \cite{CollabCoder, CoAIcoder}. Still, other systems focus on pedagogical applications, using AI to aggregate student work and support learning qualitative analysis at scale \cite{palea2024annota}. Some systems even aim to help researchers synthesize insights from their data \cite{yun2025generative}. This trend extends to commercial qualitative data analysis software, which is increasingly integrating similar AI features \cite{schroeder_llms}.

This rapid pace of tool development, however, often leaves critical tensions for reflexive methods like RTA unaddressed. A primary concern is that AI assistance can distance researchers from their data, inhibiting the deep immersion and critical questioning that is foundational to the interpretive process \cite{jiang_supporting_serendipity}. Using AI to automate interpretation also risks supplanting human agency and the critical, interpretive lens, a significant problem for reflexive analysis that often has a social justice orientation \cite{braun2019reflecting}. Furthermore, introducing opaque computational systems complicates assessments of methodological rigor and creates challenges for establishing appropriate trust and reliance on automation \cite{lee2004trust, mcdonald2019reliability}. In response, scholars have proposed that instead of seeking consensus, AI tools should be designed to highlight ambiguity and disagreement as sites for reflexive engagement \cite{jiang_supporting_serendipity, chen2018using}. Others have put forth design principles for ethical LLM integration, calling for tools that enable intentional use in specific subtasks, offer transparency, and account for researcher positionality \cite{schroeder_llms}. We build on this work by designing and evaluating a system that operationalizes these principles, focusing on an LLM-powered tool that centers researcher reflexivity and supports the communication of analytic rigor.

\subsection{Collaboration,  Reliability and Provenance in Qualitative Analysis}
Collaborative qualitative analysis involves complex coordination, requiring strategies to manage teamwork and build shared understanding, often across physical distances \cite{cornish2014collaborative, hall2005qualitative, richards2018practical, olson2000distance}. A key site of tension within this process is the concept of reliability, particularly the use of inter-coder reliability (ICR) metrics. The role and appropriateness of ICR for validating qualitative conclusions \cite{kurasaki2000intercoder} has become a subject of considerable debate \cite{o2020intercoder}. Many scholars argue that its positivist roots are fundamentally incompatible with the interpretive goals of ``Big-Q'' methods like RTA \cite{kidder1987qualitative}, a sentiment reflected in the low reporting rates of ICR in HCI and CSCW literature \cite{o2020intercoder, mcdonald2019reliability, diaz2023applying}. However, an alternative perspective reframes disagreement not as the failure to attain reliability, but as an analytically productive event that signals opportunities for deeper insight \cite{zade2018conceptualizing}. From this view, the purpose and utility of reliability metrics should be re-envisioned. Instead of serving as a definitive measure of interpretive agreement, they can be used as a diagnostic tool to spark reflexive conversations about why interpretations differ \cite{o2020intercoder}. 

In addition, a key component of managing collaboration and ensuring methodological rigor is maintaining a transparent history of how analytical concepts evolve. Prior HCI and information visualization communities have explored \emph{analytical provenance} \cite{north2011analytic, ragan2015characterizing}. While early HCI work visualized collaborative document evolution \cite{viegas2004studying, wang2015docuviz}, visualization research distinguishes between capturing low-level user actions (\emph{interaction provenance}) \cite{huang2025vistruct, cutler2020trrack} and higher-level reasoning processes (\emph{insight provenance}) \cite{gotz2009characterizing}. This distinction is critical for qualitative analysis, as interaction logs alone cannot capture the \textit{rationale} behind interpretive shifts. Research in visual sensemaking suggests that without explicit scaffolds to track these shifts, which are termed ``reframing'' in Data-Frame theory \cite{klein2007data}, analysts often struggle to recall the reasoning behind their earlier decisions \cite{lipford2010helping}. 

Currently, interpretive methods like RTA lack sufficient support for tracking the conceptual evolution of analysis, often forcing researchers to rely on labor-intensive manual versioning rather than automated provenance \cite{singh2025exploring}. While systems like CollabCoder \cite{CollabCoder} have introduced LLM-driven workflows to streamline collaboration, they explicitly prioritize rapid decision alignment. This focus on convergence risks missing opportunities to explore the roots of interpretive disagreement, leaving a critical gap in scaffolding that connects analytical choices directly back to researcher positionality.

\sysname{} is designed to address this gap. By providing an interface that automatically visualizes the evolution of codes in the analytic process and foregrounds the researchers' positionality, it structures and encourages essential reflexive conversations that are paramount to rigorous interpretive research.

\section{Formative Study}
\label{subsection:formative}

Our review of prior work revealed three primary gaps in current reflexive practices: inconsistent support for reflection during coding \cite{bowman2023using}, untracked code evolution \cite{singh2025exploring}, and a lack of visibility of the perspectives of other team members during collaboration \cite{mcdonald2019reliability}. 

To investigate these gaps in practice and identify design opportunities, we conducted a formative study with 58 qualitative researchers to understand current reflexive practices and identify design opportunities for supporting systematic reflexivity in collaborative coding environments. The study consisted of a survey targeting postgraduate researchers ($N=55$, $S_{1}$-$S_{55}$\footnote{$S_{x}$: Indicates formative \textbf{s}urvey participant number $x$}), followed by semi-structured interviews with experienced RTA practitioners ($N=3$, $I_{1}$-$I_{3}$\footnote{$I_{x}$: Indicates formative \textbf{i}nterview participant number $x$}).

\paragraph{Survey}
We conducted our survey on Prolific, targeting postgraduate researchers to understand current reflexive practices and collaborative challenges in qualitative analysis. The 30-minute preliminary survey examined how researchers engage in reflexivity during coding, track code evolution over time, and navigate differences in collaborative interpretation. We screened for familiarity with TA and prior qualitative analysis experience, retaining 55 valid responses. Participants were compensated at \$16/hour (approximately \$8 for the 30-minute survey). The protocol, consent, and compensation were approved by our university IRB office (protocol \# 49033). Details about survey questions, Prolific screening questions, and survey completion statistics can be found in Appendix \ref{appendix:formative-survey-details}.

Survey responses were analyzed using descriptive statistics for Likert-scale questions and thematic coding for open-ended responses. The first author conducted the initial analysis, and the second author reviewed the findings to ensure reliability. While the survey confirmed that researchers value reflexive practices and recognize their importance for analytical rigor, responses revealed significant implementation challenges. For example, one respondent noted: \textit{``my own assumptions can get in the way, I have to keep checking myself''} ($S_{44}$). These responses suggested that understanding the subtle workflow barriers preventing systematic reflexivity required deeper exploration than survey methods could provide.

\paragraph{Interviews}
To gain a more in-depth understanding of these implementation challenges, we conducted one-hour semi-structured interviews with three experienced RTA practitioners (2 PhD students, 1 university professor). Participants were invited through the authors' professional and academic networks. We employed purposive sampling, a non-probabilistic technique where participants are deliberately selected based on specific characteristics relevant to the research question \cite{palinkas2015purposeful}. We deemed this approach appropriate because it ensured interviewees possessed the deep RTA expertise needed to provide nuanced insights into reflexive practices that may not be evident from survey responses alone. This sampling strategy allowed us to directly verify that all practitioners met our criteria for adequate RTA experience.\footnote{All three interview participants have completed at least two formal qualitative studies as part of their research portfolio, with at least one using reflexive thematic analysis. Additionally, two of the graduate students have been formally trained in reflexive thematic analysis through a graduate-level course. The university professor has extensive experience publishing research work in reputable peer-reviewed conferences/journals in their research domain, using reflexive thematic analysis as a method. All three participants have different research foci within different sub-disciplines of HCI.} 

All interviews were conducted virtually via Zoom and facilitated by the first author. Participants received \$20 for their time. Interviews were recorded for analysis. Recorded interviews began with participants describing a recent workflow to ground the discussion in concrete practices. We then explored their approaches to reflexivity, code evolution, and collaborative interpretation. Finally, participants evaluated early system concepts and wireframes. The complete interview script is documented in Appendix \ref{appendix:formative-interview-script}.

Interview transcripts were analyzed using reflexive thematic analysis \cite{braun2019reflecting}. The first author conducted open coding, identifying patterns in how participants described their reflexive practices, code evolution strategies, and collaborative challenges. Preliminary codes and themes were discussed in weekly meetings with the research team, refining interpretations through collaborative sense-making.

\subsection{Challenges Identified from Existing Behaviors}
\label{sec:challenges}
\begin{table*}[ht]
\centering
\small
\begin{tabular}{p{2.5cm} p{5.2cm} p{3cm} p{3.0cm}}
\toprule
\textbf{Challenge (C)} & \textbf{Observed Behavior (B)} & \textbf{Design Goal (DG)} & \textbf{System Feature} \\
\midrule
\multirow{4}{=}{\textbf{C1. Inconsistent Reflection Support}} 
& \textbf{B1.} Manual reflection practices rely on individual discipline and memory 
& \multirow{4}{=}{\textbf{DG1. Systematic In-the-Moment Reflexivity Support}} 
& \multirow{2}{=}{(A) \reflexivecolor{In-situ reflexive exercises}} \\
\cmidrule(l){2-2}
& \textbf{B2.} Insight recording happens in separate documents disconnected from codes 
& 
& \\
\midrule
\multirow{4}{=}{\textbf{C2. Untracked Code Evolution}} 
& \textbf{B3.} Version control documentation is labor-intensive and inconsistent 
& \multirow{4}{=}{\textbf{DG2. Automated Code Evolution Tracking}} 
& (B) \transparentcolor{Analytical history}\\
\cmidrule(l){2-2} \cmidrule(l){4-4}
& \textbf{B4.} Reactive change management relies on researcher awareness to trigger documentation
& 
& (C) \transparentcolor{Code drift alert}\\
\midrule
\multirow{6}{=}{\textbf{C3. Inaccessible Collaboration Perspectives}} 
& \textbf{B5.} Information gathering relies on voluntary self-disclosure in meetings 
& \multirow{6}{=}{\textbf{DG3. Collaborative Positionality Integration}} 
& (D) \collaborationcolor{Positionality-aware discussion prompts} \\
\cmidrule(l){2-2} \cmidrule(l){4-4}
& \textbf{B6.} Structured discussion happens separately from actual coding decisions 
& 
& (E) \collaborationcolor{Discussion focus for disagreements} \\
\cmidrule(l){2-2} \cmidrule(l){4-4}
& \textbf{B7.} Proactive reflective prompting occurs informally and sporadically 
& 
&\\
\bottomrule
\end{tabular}
\caption{Mapping from user challenges and behaviors to design goals and system features.}
\label{tab:challenge_strategy_goal}
\end{table*}

Our survey and interview data reveal that researchers actively engage in reflexive practices but struggle with tools that do not support the non-linear nature of RTA. Table \ref{appendix:formative-survey-details} maps these observed behaviors to design goals and system features.

\subsubsection{Challenge 1: Inconsistent Reflection Support}
\label{sec:challenge1}
Our survey revealed that qualitative researchers desire deep reflexivity, but implementation is sporadic. While respondents ($40/55$) ``often'' or ``always'' considered how their experience shaped interpretation, they lacked integrated tools to capture it. Researchers rated ``asking myself guiding questions'' as highly helpful, yet currently rely on disconnected methods like ``writing a separate research journal'' ($32/55$). As $S_{44}$ noted, ``my own assumptions can get in the way,'' yet without integrated support, checking these assumptions became a disruption.

\paragraph{Design Considerations} To address this fragmentation, participants suggested distinct alternatives. $I_1$ advocated for a ``spatial whiteboarding'' interface similar to LiquidText \cite{LiquidText} to visually map connections, while $I_3$ suggested ``stage-based checklists'' enforced before and after coding sessions. However, we reasoned that spatial tools often separated analysis from the raw text, and stage-based checks missed the immediate moment of interpretive decision-making. Consequently, we prioritized \textit{in-the-moment reflexivity support} (DG1) to keep the raw text visible. We integrated $I_2$'s suggestion for a \textit{user-initiated reflection}, which allowed researchers to engage in reflection at their own pace with self-selected reflection areas, rather than time-based or progress-based triggers.

\subsubsection{Challenge 2: Untracked Code Evolution}
\label{sec:challenge2}
Although researchers recognized the importance of tracking how codes shift, current approaches were manual and reactive. To manage code drift, researchers updated definitions ($37/55$) or split codes ($31/55$), often relying on memory ($29/55$) to maintain consistency. $S_{17}$ described maintaining a ``living codebook,'' while $S_3$ used a ``separate tab for the initial code book'' to duplicate data.

\paragraph{Design Considerations} Participants emphasized the need to control how code definitions change. $I_2$ suggested implementing hierarchical role-based visibility, where a ``project lead'' blinds junior researchers to specific codes to ensure independent thought. We addressed this by incorporating flexible visibility controls that allow users to selectively display or hide analysis progress. However, we reasoned that while hiding information could reduce bias, maintaining rigor required the ability to audit how those definitions shifted over time. Therefore, we prioritized \textit{transparent provenance} (DG2) as the underlying framework, adopting $I_1$'s suggestion to treat code history as an exportable artifact useful for open science and pedagogy.

\subsubsection{Challenge 3: Inaccessible Collaboration Perspectives}
\label{sec:challenge3}
Accessing and integrating collaborator positionalities remained a persistent friction point. While respondents valued understanding a partner's ``disciplinary perspective'' ($35/55$) and ``relevant lived experiences'' ($37/55$), this information was rarely integrated into the coding interface. $S_{34}$ highlighted how ``communication breakdowns'' occur when trying to ensure consistency while respecting diverse perspectives. Currently, teams rely on informal discussions to surface these lenses, which $I_1$ noted led to inconsistent awareness of ``why'' interpretations differ.

\paragraph{Design Considerations} Participants proposed specific reflexive interventions, such as ``fairness audits'' ($I_3$) or ``dominant narrative'' critiques ($I_1$), where researchers explicitly examined how their findings fit within or resist the prevailing discourse of the field. We reasoned that hard-coding these specific theoretical frameworks might limit the tool's applicability across different RTA contexts. Instead, we focused on the structural mechanism of ``double coding'' proposed by $I_3$, which treated conflicting interpretations as valid and parallel data points rather than errors to be resolved. We aimed to surface \textit{principled disagreement} (DG3) by displaying these divergent codes side-by-side with researcher profiles. This facilitated the ``why not'' analysis ($I_3$), in which users considered why they \textit{missed} a perspective held by their partner.

\subsection{Design Goals}
\label{sec:design-goals}
Drawing from our formative study's observed challenges and design considerations, we formulated three design goals to enhance systematic reflexivity while preserving researcher agency:

\begin{enumerate}
    \item \textbf{Systematic In-the-Moment Reflexivity Support (DG1):} To address Inconsistent Reflection Support (C1), which arose from inconsistent manual reflection and fragmented insight recording (B1, B2), we aimed to provide structured, in-the-moment reflexive prompts. This goal aimed to integrate reflection seamlessly into the coding workflow, anchoring insights to specific data without requiring researchers to switch contexts.
    \item \textbf{Automated Code Evolution Tracking (DG2):} To address Untracked Code Evolution (C2), originating from labor-intensive documentation (B3) and reactive change management (B4), our objective was to automatically create a transparent, auditable history of code evolution. This includes proactive Code Drift Alert to move beyond the reliance on manual researcher awareness (B4), making the entire analytical journey, including splits, merges, and definition edits, both visible and rigorous.
    \item \textbf{Collaborative Positionality Integration (DG3):} To address Inaccessible Collaboration Perspectives (C3), a challenge stemming from reliance on informal self-disclosure (B5), separate discussions (B6), and sporadic prompting (B7), we aimed to make positionality an actionable and visible part of the analysis by surfacing interpretive differences and scaffolding dialogue around researchers' unique lenses.
\end{enumerate}
\section{System Design — \sysname{}}
\label{sec:system}

With the design goals in \S\ref{sec:design-goals} (DG1–DG3) in mind, we implemented \sysname{}, a web-based collaborative environment for reflexive thematic analysis that centers in-situ reflection, records code evolution for transparency, and makes positionality actionable. We first outline a usage scenario, then detail the core features, the prompt design that operationalizes our reflexive practices, and finally, technical details. We open-sourced the code for \sysname{} at: \href{https://github.com/harryye930/reflexis}{https://github.com/harryye930/reflexis}.

\subsection{Workflow \& Usage Scenario}
\label{sec:workflow}
To demonstrate \sysname{}'s integrated workflow, we follow two analysts, Alice and Bob, as they move from project setup to final reporting.

\paragraph{Step 0: Profile and project setup.}

Alice, a seasoned policy analyst in municipal government, initiates a project with Bob, a university researcher specializing in social equity. They begin by completing brief researcher profiles on \sysname{}, communicating their respective disciplinary lenses and lived experiences to ground the system's future interactions.

\paragraph{Step 1: Independent RTA Coding with Just-in-Time Reflection. }

During independent coding, Alice encounters a city manager's quotation on feedback systems. Activating the \reflexivecolor{\emph{ReflexiveLens}}, she uses the \reflexivefeature{Positionality} prompt to note that her government background reveals \textit{``a deep-seated institutional fear of failure,''} and codes it \code{Bureaucratic Hurdles}. Bob, coding the same passage through his academic lens, notes that the system \textit{``privileges those with resources,''} applying \code{Participation Barrier}.

Later, as Alice reuses her code for a budget issue, the \transparentcolor{\emph{Code Drift Alert}} flags a semantic shift from ``process'' to ``finance''. Agreeing with the alert, she utilizes the \transparentfeature{Split} option to create a new, more precise code: \code{Financial Constraints}.

\paragraph{Step 2: Team Review Guided by Positionality and Attention Routing.}

Transitioning to the shared workspace, the pair uses the \collaborationcolor{\emph{Discussion Focus}}'s \collaborationfeature{Show Diff View} to isolate divergent coding. On their first diverging analysis, a \collaborationcolor{\emph{Positionality-aware Discussion Prompt}} immediately bridges their profiles: 

\begin{quote}
\textit{
\collaborationfeature{Alice's government lens sees an implementation} 
\collaborationfeature{issue, while Bob's academic lens sees an equity} 
\collaborationfeature{issue. How do these perspectives reveal different}
\collaborationfeature{aspects of the problem?}}
\end{quote}

This contextualization transforms a coding conflict into a productive dialogue. Realizing neither code suffices, and after reviewing the \transparentcolor{\emph{Code-level Analysis History}} to track down Bob's prior usage of that code, they merge their initial attempts into a new code: \code{Systemic Exclusion}.

\paragraph{Step 3: Reporting and Transparency.}

Finally, to write their methodology section, they consult the \transparentcolor{\emph{Project-level Analysis History}}. This visual narrative traces the exact provenance of \\ \code{Systemic Exclusion}, allowing them to export an audit trail that substantiates the rigor and evolution of their collaborative analysis.

\subsection{Key Features}
\label{sec:features}
\subsubsection{\reflexivecolor{ReflexiveLens: In-situ reflection and situated viewing}}
\begin{figure*}[!ht]
    \centering
    \includegraphics[width=.8\linewidth]{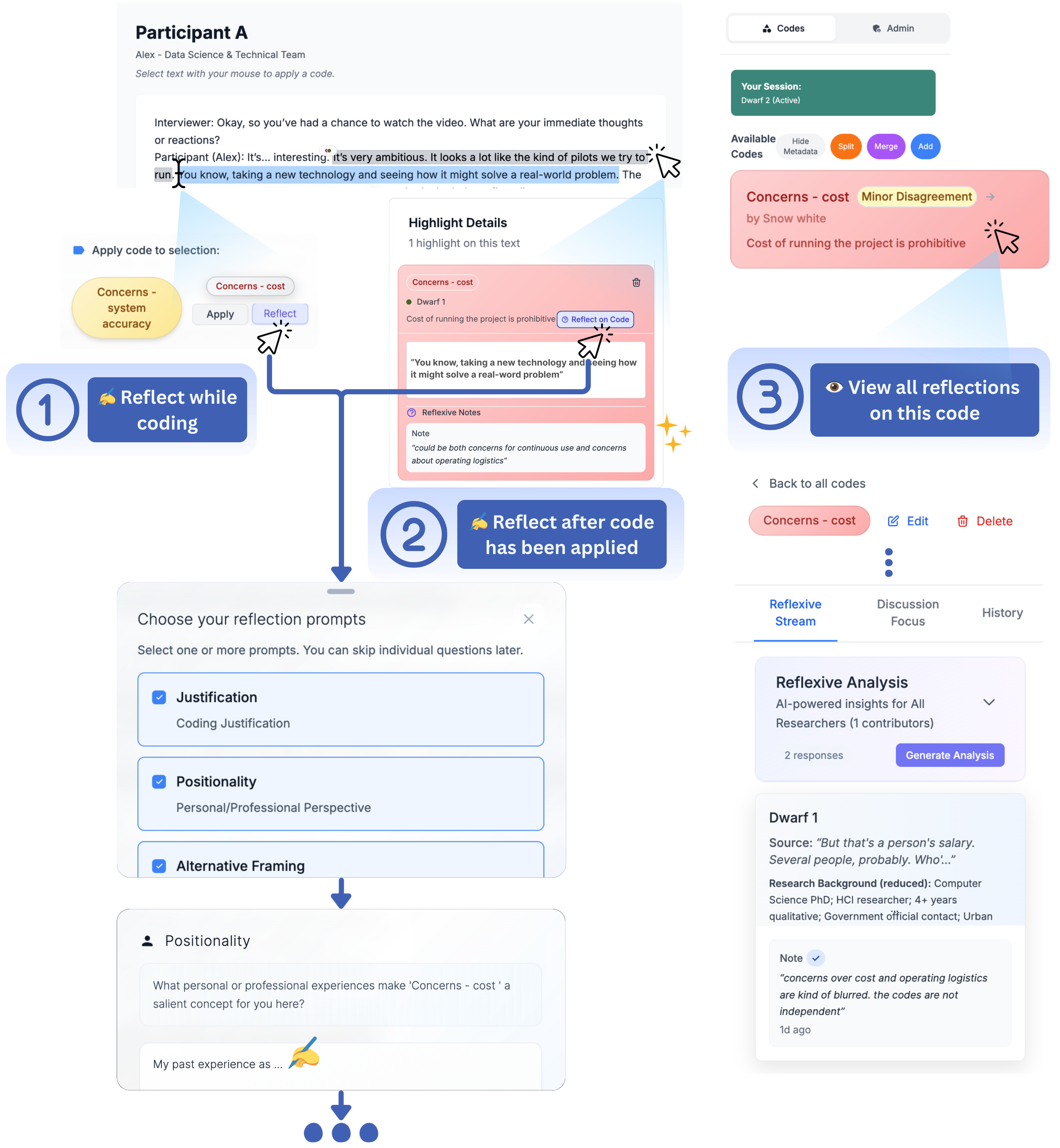}
    \caption{\sysname{}: \reflexivecolor{Supporting Reflexive Analysis}. \reflexivecolor{(1)} Users can easily engage in a just-in-time guided reflection during analysis. The reflection process is designed to be lightweight and optional, reducing labor while encouraging reflexivity. All users can review reflexive notes directly under the coding workspace, or \reflexivecolor{(2)} engage in additional reflection. \reflexivecolor{(3)} Users can also access reflexive notes for a specific code within the sidebar and filter by unique user, in addition to creating structured summaries.}
    \label{fig:system-reflexive-lens}
\end{figure*}

To support \hyperref[sec:dg1]{DG1}, \sysname{} promotes and surfaces reflexivity during key analytic moments. When a coder applies a code to a text snippet, or reflects on any prior analysis, the coder can use the \reflexivecolor{\emph{ReflexiveLens}} panel to add reflexive memos. Drawing from our formative study's findings on inconsistent reflection support (see Section \ref{sec:challenge1}) and established reflexive practices (e.g., \citet{probst2014double}), the panel provides a series of pre-populated prompts to help anchor the coder's analysis. These include prompts for: 
\begin{enumerate}
    \item \reflexivefeature{justification of the code just applied};
    \item \reflexivefeature{how coder’s background makes this interpretation}
    \reflexivefeature{salient};
    \item \reflexivefeature{other plausible alternatives}.
\end{enumerate}
Complementing these structured prompts, a free-form \reflexivefeature{notes} section is included to capture any emergent insights or unstructured reflections that fall outside the provided categories. The interaction is deliberately lightweight and optional, so that reflection occurs ``just-in-time,'' without interrupting the coding process.

Beyond capturing reflective thoughts, \sysname{} provides ways to connect them back to the corpus. Users can see an overlay that renders reflexive notes directly below the coded excerpts, allowing coders to re-encounter prior rationales as they move through adjacent material and to append follow-up reflections in place. A \reflexivecolor{\emph{reflexive stream}} aggregates these entries across documents as a time-ordered, filterable feed (by code or coder), enabling teams to trace how interpretations consolidate, shift, or bifurcate over repeated encounters with the data. By situating reflexive notes alongside textual evidence, \sysname{} treats them as first-class, inspectable artifacts. This approach reframes reflexivity from an after-the-fact supplement to a continuous RTA practice. 
See Figure \ref{fig:system-reflexive-lens}.

\subsubsection{\transparentcolor{Making Code Evolution Transparent and Tangible}}
\begin{figure*}[!ht]
    \centering
    \includegraphics[width=.7\linewidth]{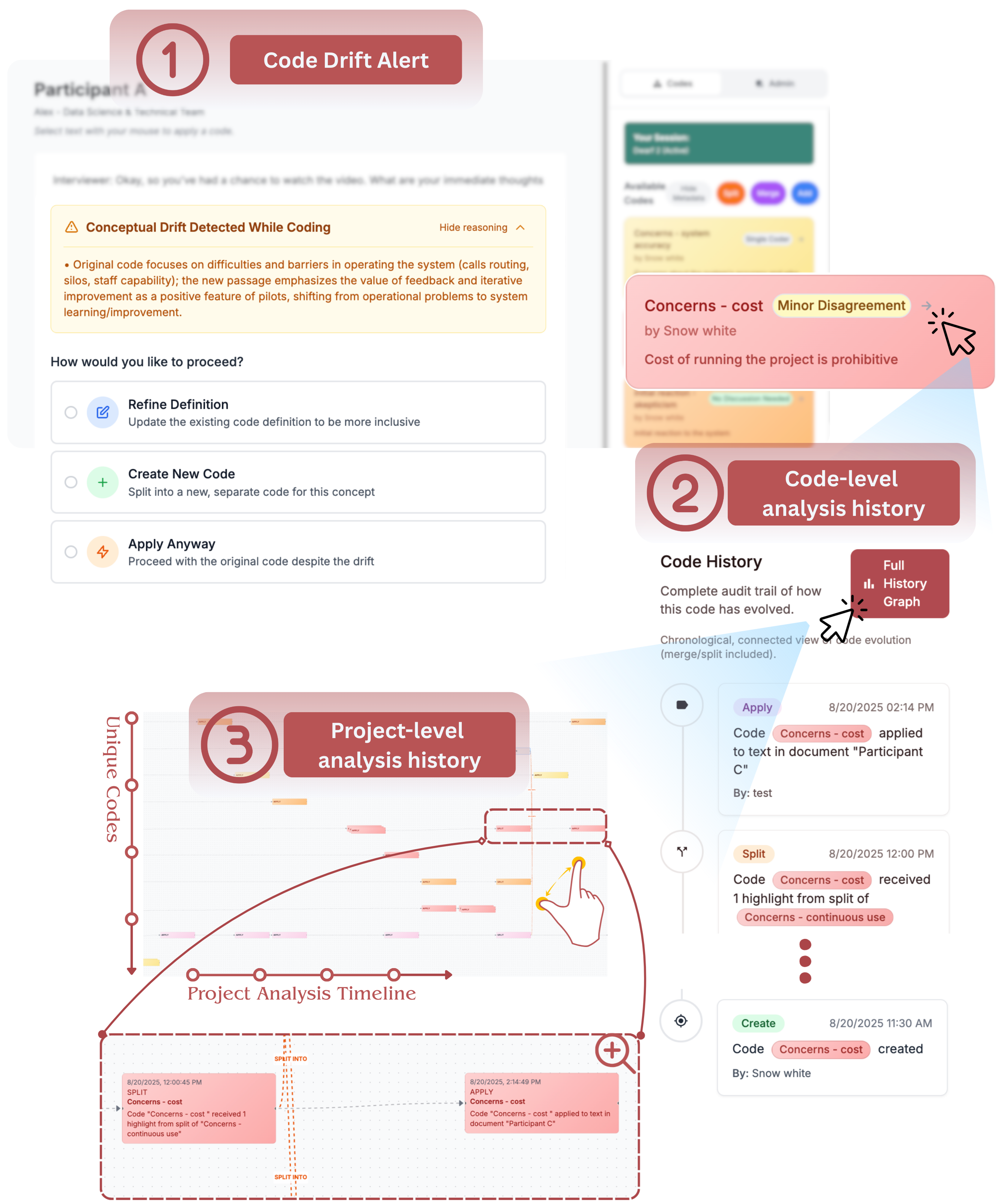}
    \caption{\sysname{}: \transparentcolor{Making Code Evolution Transparent and Tangible}. To make the analytical process visible and consistent, \sysname{} first \transparentcolor{(1)} actively engages users with codes as evolving concepts by monitoring for code drift. When a shift in code application is detected, it alerts researchers, prompting a deliberate re-evaluation of the ongoing analysis and encouraging reflective practices. To support this reflective process, the system provides a detailed record of the analysis: the \transparentcolor{(2)} code-level analysis history records every analytic event, including code creation, application, merges, and splits. Finally, \transparentcolor{(3)} the project-level analysis history visualizes analytical provenance, tracing how codes are transformed over time.}
    \label{fig:system-code-evolution}
\end{figure*}

To support \hyperref[sec:dg2]{DG2}, \sysname{} treats the codebook as a living object and records its evolution at two levels. At the code level, the system maintains a versioned \transparentcolor{\emph{code-level analysis history}} including \transparentfeature{creation}, \transparentfeature{renames}, \transparentfeature{definition edits}, \transparentfeature{merges}, and \transparentfeature{splits}, with each user action timestamped and attributed to the user who executed it. This enables readers to understand not only \emph{what} changed but also \emph{how} the code came to be. At the project level, a global \transparentcolor{\emph{project-level analysis history}} aggregates events across codes and documents to support coordination (i.e., ``what changed since our last meeting?'') and reporting (i.e., ``how did the codebook evolve since the beginning of the project?''). Coders zoom in on the timeline to reveal details-on-demand, and they can export snapshots for auditing or producing methods and appendices in their final reports.

To help researchers engage with this evolution, the LLM-enabled \transparentcolor{\emph{Code Drift Alert}} monitors each new application of a code against an evolving set of exemplars of the code provided by the users, flagging potential drift (e.g., broadening in definition or misuse of a code). A coder can then choose to \transparentfeature{refine the definition} of the existing code, \transparentfeature{split it into a new code}, or consciously override the alert and \transparentfeature{apply the original code}. We hypothesize that these mechanisms make code evolution inspectable throughout the study, reduce misunderstandings in multi-coder teams, and provide concrete materials to promote transparency in the final report. See Figure \ref{fig:system-code-evolution}.

\subsubsection{\collaborationcolor{Scaffolding Collaborative Interpretation}}
\begin{figure*}[!ht]
    \centering
    \includegraphics[width=\linewidth]{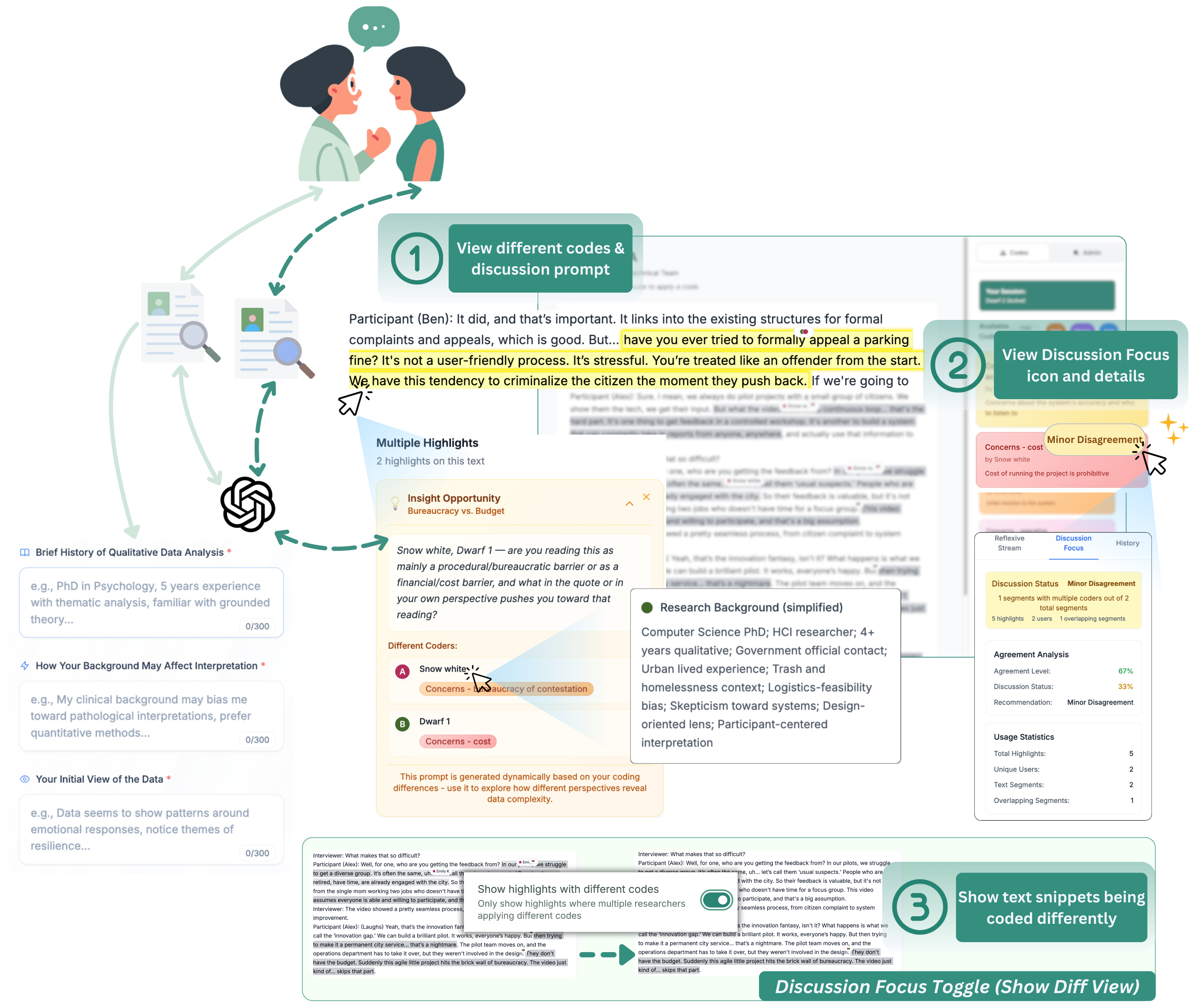}
    \caption{\sysname{}: \collaborationcolor{Scaffolding Collaborative Interpretation}. To ground collaboration in reflexive practice, researchers first complete profiles detailing their background and positionality. When interpretive differences arise during coding, \collaborationcolor{(1)} \sysname{} surfaces them through a mediated discussion prompt that explicitly references the researchers' stated positionalities, helping to contextualize their unique lenses and scaffold constructive disagreement. \collaborationcolor{(2)} To help teams manage and prioritize these conversations, a Discussion Focus icon is displayed next to each code, providing an at-a-glance overview of agreement levels and a detailed breakdown after click-through. Furthermore, \collaborationcolor{(3)} researchers can toggle the view to show only data snippets with divergent codes, focusing the team's attention on areas requiring discussion.}
    \label{fig:system-collaborative-interpretation}
\end{figure*}

To support \hyperref[sec:dg3]{DG3}, \sysname{} makes positionality actionable when collaborators come together. For any coded passage, highlights and codes are co-presented, and the system issues \collaborationcolor{\emph{Positionality-Aware Discussion Prompts}} that explicitly connect divergent readings to coders’ stated positionalities. Collaborators can discuss not only ``what code did you use?'' but also ``why did this reading make sense to you, given your background?'', maintaining a principled divergence that advances the analysis. Our implementation operationalizes positionality through concise, structured, self-authored researcher statements. While this can help teams recognize that disagreement is rooted in situated lenses rather than individual error, it also risks over-asserting these links, i.e., implying that a given interpretation is determined by a researcher's role or identity. We therefore view the prompt as an optional scaffold that surfaces one possible explanation for disagreement, encouraging further discussion, rather than as a definitive attribution of why a coder read the data in a particular way.

The \collaborationcolor{\emph{Discussion Focus}} feature routes user attention toward areas of potential interpretive divergence using two coordinated components. First, a \collaborationfeature{discussion focus icon} appears beside individual codes that may be too broad (applied to many different contexts) or where there is potential disagreement (used in unique ways by different coders). Separately, researchers can also toggle the \collaborationfeature{show diff view} to display only the data snippets being coded differently by multiple researchers. Together, these mechanisms help a team to better spend limited meeting time engaging in deeper, targeted discussions. By design, the discussion focus centers analytic attention without enforcing convergence; teams may choose principled divergence if appropriate. All collaborative work is synchronized in real-time, allowing multiple coders to work on the same corpus concurrently. See Figure \ref{fig:system-collaborative-interpretation}.

\subsection{Prompt Design and LLM Usage}
\label{sec:llm}

We adopted a narrow, specific role for language models in \sysname{}: assist reflection and discussion without generating code or suggesting what to apply. All prompt wordings and safety constraints are provided in Appendix \ref{appendix:llm-prompt}; only function and scope are described here.

Models are used in four features. A researcher can generate a digest of linguistic patterns, positionality effects, and articulated alternatives with \reflexivecolor{\emph{reflexive stream structured summary}}. This allows researchers to gain a high-level overview of their team's analytical approach and helps to surface emergent insights for further development. For \collaborationcolor{\emph{Positionality-Aware Discussion Prompts}}, short discussion prompts are synthesized from the coded context and coders’ positionality profiles to help collaborators explain why interpretations may differ. The system does not adjudicate or prescribe one researcher's choices over another. For \transparentcolor{\emph{Code Drift Alert}}, the system compares the user-selected candidate corpus with recent exemplars from the same code, and flags potential broadening or misuse for human review. Finally, \collaborationcolor{\emph{Positionality Keywords}} provide concise, LLM-simplified cues from a coder’s profile and recent memos, shown beside their highlights to keep backgrounds salient during discussion. \sysname{} uses OpenAI's \texttt{GPT-5} and \texttt{GPT-5-mini} with \texttt{low} to \texttt{medium} reasoning effort.

\subsection{System Implementation}
\label{sec:implementation}
The frontend was built with Next.js and Tailwind CSS, with the main coding workspace supporting character-precise highlighting and overlapping annotations. Project-level analysis history was visualized using the React Flow library.

We used Google's Firebase services for real-time collaboration. In particular, the Firestore database was used for real-time synchronization, anonymous authentication for easy onboarding and user identification, and managed hosting for backend deployment. The UI subscribed to shared documents, annotations, and profiles, so updates appeared instantly across clients without manual refresh. To maintain transparency, every analytic action (apply, edit, rename, merge, split) was written to an immutable event log that provides history views and exports.

\subsubsection{Percentage Agreement}
The agreement metric for \emph{Discussion Focus} is calculated based on percentage agreement, a common metric in qualitative analysis \cite{miles1994qualitative}, using the following formula: 
\newpage 
\begin{equation*}
    Agreement = \frac{\text{Count of highlights agreed by at least one other coder}}{\text{Total count of highlights contributed by all coders}}
\end{equation*}

We used this metric because common alternatives like Cohen's Kappa can be statistically unsound when applied to more than two coders due to the violation of statistical assumptions \cite{mchugh2012interrater}. We emphasize that this metric (and its color cue) is used solely to route attention to potentially fruitful discussion points, not to optimize for high agreement, consistent with our stance that principled divergence is often desirable in RTA. Yet, we also note the possibility that visual encodings may nudge behavior toward consensus \cite{leshed2010visualizing}.
\section{User Evaluation}

To evaluate \sysname{} and answer our research questions, we ran a single-condition study to examine how \sysname{} supports reflexive thematic analysis (RTA) in collaborative settings.

\subsection{Participants, Setting, and Ethics} 
We recruited 12 participants ($P_1$-$P_{12}$) via social media (i.e., LinkedIn, Twitter) and other public or private Slack workspaces. The inclusion criteria were researchers who are experienced in qualitative data analysis and familiar with RTA concepts and practices (i.e., reflection, positionality, etc). To ensure a high level of proficiency, we screened prospective participants by requesting a summary of their past experience with RTA. Specifically, we asked candidates to detail relevant projects and possible peer-reviewed publications that utilize RTA. Only those demonstrating sufficient expertise (i.e., who had worked on at least one RTA project prior to the study) were invited to participate in the study.

The majority of participants (10/12) were graduate students (i.e., master's, PhD), one participant was a university professor, and another worked as a data analyst. Most of the participants came from an HCI background (10/12), with intersections spanning across NLP, visualization, and computing education. The rest of the participants had a more technical background (i.e., software engineers, 2/12). On average, participants had 3.75 years of qualitative data analysis experience (Min: 1, Max: 7, Median: 4). All participants described Reflexive Thematic Analysis (RTA) as one of the methods they used to analyze qualitative data in the past. Participants reported mainly using spreadsheet software (i.e., Microsoft Excel, Google Sheets; 10/12) and text editors (i.e., Microsoft Word, Google Docs; 8/12) as part of their past RTA projects, with a handful of participants choosing QDA software like NVivo (5/12), ATLAS.ti (2/12), and Dedoose (1/12). Participants did not have a consensus for norms on reporting about their analytical practices or artifacts for the broader scientific community. Half of the participants indicated they did not share any artifacts, while three indicated they would share an end-of-study statement. Participants were asked to share their theoretical background and view on the topic of analysis (discussed below) without seeing the actual corpus, and were paired up to prioritize diverse perspectives, leading to the formation of 6 pairs (see Appendix \ref{appendix:demographics}). 

All sessions took place remotely. Each session lasted approximately 1.5-2 hours, and participants received compensation at a rate of \$20/hour upon successful completion. The protocol, consent, and compensation were approved by our university IRB office (protocol \# 49033).

\subsection{Study Design}
In this study, participants were instructed to use \sysname{} to perform reflexive thematic analysis (RTA) as a pair. Our research goal was to understand how \sysname{} might reshape the deep, interpretive practices of collaborative qualitative analysis, as opposed to measuring efficiency gains. Consequently, a direct comparison to a baseline tool would not effectively serve our research objective. We further explain our reasoning for not employing a standard baseline condition in Section \ref{sec:no-baseline-justification}.

Our study protocol on collaborative reflexive analysis followed common qualitative guidance emphasizing (i) independent analytic passes, (ii) transparency and evolution of code definitions, and (iii) negotiated understanding via discussion. In particular, we referred to the process documented by \citet{o2020intercoder} (see Figure 3 in the corresponding paper). We randomly assigned one participant as Coder 1 and the other as Coder 2. Coder 1 first provided the initial code definitions and completed an initial pass on the provided corpus. Coder 2 then coded the same snippets using Coder 1's generated code list, but without seeing which codes Coder 1 had applied to text segments. The pair subsequently reviewed differences and rationales, mediated by system features that surface code overlaps, definition drift, and unmatched assignments.

Following prior evaluations of AI-supported qualitative coding tools that adopt intensive, single-session designs to study collaborative sense-making in depth (e.g., \cite{PaTAT, CollabCoder}), our study similarly invites pairs to complete a bounded, collaborative RTA episode with \sysname{}. While a 1.5-hour session cannot fully capture a multi-week qualitative analysis practice, it enables us to closely observe how analysts interact with and reason through \sysname{}’s reflexivity-oriented features within a realistic, collaborative coding session.

\subsubsection{Reasoning for Not Employing a Baseline Condition}
\label{sec:no-baseline-justification}
RTA is intentionally non-prescriptive regarding specific tools or workflows. The method is positioned as an ``adventure, not a recipe'' as described by \citet{braun2019reflecting}, prioritizing theoretical flexibility and a recursive analytical process over rigid, linear steps \cite{braun2006using}. Reflecting this methodological freedom, RTA practitioners commonly employ a diverse and often bespoke combination of tools. Workflows can rely on physical notes \cite{probst2014double}, leverage general-purpose software like Microsoft Word, which is often deemed sufficient \cite{clarke2021practical}, or incorporate dedicated qualitative analysis software \cite{silver2014using}. As there is no single, standard RTA toolchain, any attempt to impose one as a baseline would create an artificial scenario, confounding the study with the cognitive overhead of learning a new baseline tool, and hindering our ability to yield meaningful insights into real-world analytical practice.

Instead, our study was designed to observe how participants integrated \sysname{} into their workflows, allowing us to gather rich, situated insights into its impact on their analytical process. This approach enables us to evaluate our system on its own terms via its ability to support the core principles of reflexive analysis \cite{braun2021one}.

\subsubsection{Analysis Dataset} 
The analysis corpus used during the evaluation study was a synthetic, context-faithful set of interview transcripts derived from a prior CHI paper on ``Contestable Camera Cars'' \cite{camera-car}. We chose this research topic because the original study employed reflexive thematic analysis, aligning with our methodology. In addition, the subject of ``Contestable Camera Cars'' is readily understandable and requires minimal domain-specific knowledge, making it highly accessible for interpretation. We prompted a large language model\footnote{Given identical system prompts, we compared \texttt{gemini-2.5-pro} (thinking budget: 32k tokens) and \texttt{gpt-5} (reasoning effort: high). We collectively elected to use the \texttt{gemini-2.5-pro} outputs, which yielded more realistic, conversational transcripts for our study.} with the reference paper to generate three distinct interview transcripts that (i) preserved the tone and terminology while staying within the civic-technology context; and (ii) echo the themes reported by the original authors. Participants were not informed that the transcripts were synthetic, and were instructed to analyze one of the three synthetic transcripts as a pair, in random order. This choice preserved ecological validity by retaining content and findings from a real study while avoiding privacy concerns and domain prerequisites. The full system prompt is available in Appendix \ref{appendix:llm-prompt-user-study}.

\subsection{Procedure}
The study sessions were conducted remotely and followed a structured protocol consisting of four distinct phases: setup and orientation, independent coding, mediated negotiation, and post-task reflection. Prior to the start of the session, the moderator loaded the segmented transcript corpus and initialized an empty project within \sysname{}. After confirming participant consent and initiating audio and screen recording, the moderator instructed participants on the think-aloud protocol and the norms for respectful, reflexive discussion.

The session began with a 15-minute orientation. The moderator framed the upcoming activity as a typical RTA pass and demonstrated the core features of \sysname{}. Participants were briefed on the study’s specific collaborative workflow, which required sequential roles (Coder 1 followed by Coder 2) prior to a joint discussion, following the process described by \citet{o2020intercoder}.

Following orientation, participants engaged in an independent coding phase lasting approximately 30 minutes. Coder 1 worked first (approx. 15 minutes) to code assigned snippets, creating or renaming codes as necessary, writing short reflexive notes, and refining code definitions. Coder 2 then coded the same snippets (approx. 10 minutes) utilizing the code list and definitions developed by Coder 1. Furthermore, while Coder 2 utilized Coder 1's schema, they were blinded to Coder 1’s specific text segment code assignments to ensure independent application.

Once independent passes were complete, the pair transitioned to a 20-minute phase of mediated comparison and discussion. \sysname{} surfaced provisional differences between the coders, such as unassigned snippets, definition drifts, or competing codes. The pair reviewed these discrepancies alongside their reflexive notes to explain their rationales and negotiate updates to code meanings. This phase concluded with a brief, shared reflection on how their respective positionalities and tool support influenced their analytical decisions.

Finally, the session closed with a 20-minute evaluation block consisting of a short post-study survey and a semi-structured interview. These instruments were designed to capture participants' perceptions of the system's usefulness, its compatibility with existing qualitative workflows, and the specific collaboration dynamics enabled by \sysname{}.

\subsection{Measures and Analysis} 
We focused on qualitative feedback and perceived usefulness gathered through (i) brief pre/post-study surveys (open-ended items on reflexivity, fit with existing workflow, and perceived support), (ii) a semi-structured post-study interview, and (iii) in-study interactions (think-aloud comments, on-canvas/reflection notes, and mediated discussion logs). We also captured descriptive traces of the analysis process via screen recording, e.g., codebook edits (merges/splits/renames), definition revisions, and disagreement flags, to contextualize participants’ narratives.

We did not measure task time, ``accuracy'' of any kind (i.e., inter-rater reliability, participants' accuracy against `ground truth'), or other aggregate quantitative metrics. In RTA, higher agreement or faster completion does not necessarily indicate better analytic quality; as such metrics can inadvertently suppress productive interpretive differences and reflexive engagement \cite{mcdonald2019reliability}. Accordingly, our measurements were selected to illuminate how interpretations evolve and are negotiated, not to compute efficiency or reliability scores.

\subsubsection{Qualitative Data} We conducted RTA for qualitative data including in-study reflections and interview transcripts \cite{braun2019reflecting}. The analysis was led by a graduate student formally trained in qualitative methods, with prior research experience across HCI domains, including human-centered AI, computing education, and visualization. The process was collaborative, involving iterative discussions on codes and themes with a broader research team that included two faculty supervisors and other HCI and visualization researchers. As part of our reflexive practice, we continually questioned how our backgrounds might shape our interpretations. For instance, as designers and researchers of interactive systems, we asked: How do our own writing and feedback-seeking practices influence what we notice in participants' accounts?  While these reflexive discussions do not eliminate subjectivity, they were essential for ensuring our workflow remained an interpretive and dialogic process.

\subsubsection{Quantitative Data} To supplement our primary qualitative data from interviews and observations, we administered short surveys both before and immediately after the study. The pre-study survey included questions on demographics and individuals' self-reported reflection practices using their own habitual QDA tools. The post-study survey included Likert-scale items assessing the experience and usability of the tool, as well as questions about the frequency of reflection using \sysname{}. For the frequency question, we paired the responses from each participant from the pre-study survey and post-study survey to examine any differences in analysis practices. 

In designing the Likert items, we constructed a composite instrument designed to capture dimensions specific to collaborative RTA that are not covered by generic standardized questionnaires. To ensure construct validity, we adapted items from established instruments. Specifically, we derived items regarding usability and workload (Q20, Q21, Q23) from the System Usability Scale (SUS) \cite{lewis2018system} and NASA-TLX \cite{hart2006nasa}; items regarding transparency, agency, and trust (Q8--12, Q18) from XAI evaluation frameworks \cite{hoffman2023measures}; and items regarding collaboration and safety (Q15--17) from literature on psychological safety \cite{edmondson1999psychological}. These were combined with domain-specific process checks (e.g., Q1--7, Q22, Q25--31) derived from Braun and Clarke's criteria for quality in RTA \cite{braun2021one}. We report these scores descriptively to contextualize the qualitative feedback.

\subsubsection{LLM Technical Analysis} 
While the ultimate utility of an interpretation in RTA is determined by the researcher, we sought to verify the general quality of the output from LLM-enabled features through a two-stage process: a preliminary quality check during the pilot study and a post-hoc technical analysis. During the pilot phase, we iteratively tested and refined the system prompts to ensure baseline coherence. Following the evaluation study, two co-authors performed a post-hoc analysis of the interaction logs, reviewing every occurrence of Code Drift Alert ($N=7$) and Positionality-aware Discussion Prompt ($N=18$) as they naturally triggered during the study.

We evaluated these outputs based on \textit{relevance}, specifically verifying that the content was grounded in the provided context rather than containing hallucinations (e.g., detecting drift when there is none, or referencing non-existent text segments). Outputs were binary coded as \textit{Relevant} (logically derived from the available context, code definitions, and researcher background) or \textit{Irrelevant} (factually disconnected from the input). The two raters achieved perfect consensus. We chose relevance over ``correctness'' or ``accuracy'' since in an interpretivist paradigm, whether a shift in meaning constitutes ``drift,'' or whether a positionality connection is ``true'' is not an objective state for the system to predict, but a subjective judgment for the analyst themselves \cite{braun2019reflecting}. These dynamics are detailed in the qualitative results below.

\subsection{Rationale and Limitations}
A single-condition design privileged depth of interpretive work over tool-to-tool comparisons. This limits claims about comparative effectiveness but aligns with RTA’s interpretivist commitments and avoids incorporating non-standard ``baselines''~ \cite{braun2023toward, carminati2018generalizability}, as discussed in Section \ref{sec:no-baseline-justification}. We partially mitigated the lack of a baseline by (a) eliciting participants’ customary tool use and practices, and (b) observing disagreement-focused collaboration during the study session. Using an LLM-generated dataset from a real-world example preserves the original civic context while minimizing sensitivity risks \cite{tsai2016promises}; however, synthetic data may narrow variability relative to field data and could subtly steer theme salience.

\section{User Study Results}

\subsection{Reframing Reflexivity: \sysname{} Encouraged Granular, In-Situ Reflection Amidst Workflow Tensions \reflexivecolor{(RQ1)}}
\label{sec:result-reflexive-1}

\begin{figure}[ht]
    \centering
    \includegraphics[width=.9\linewidth]{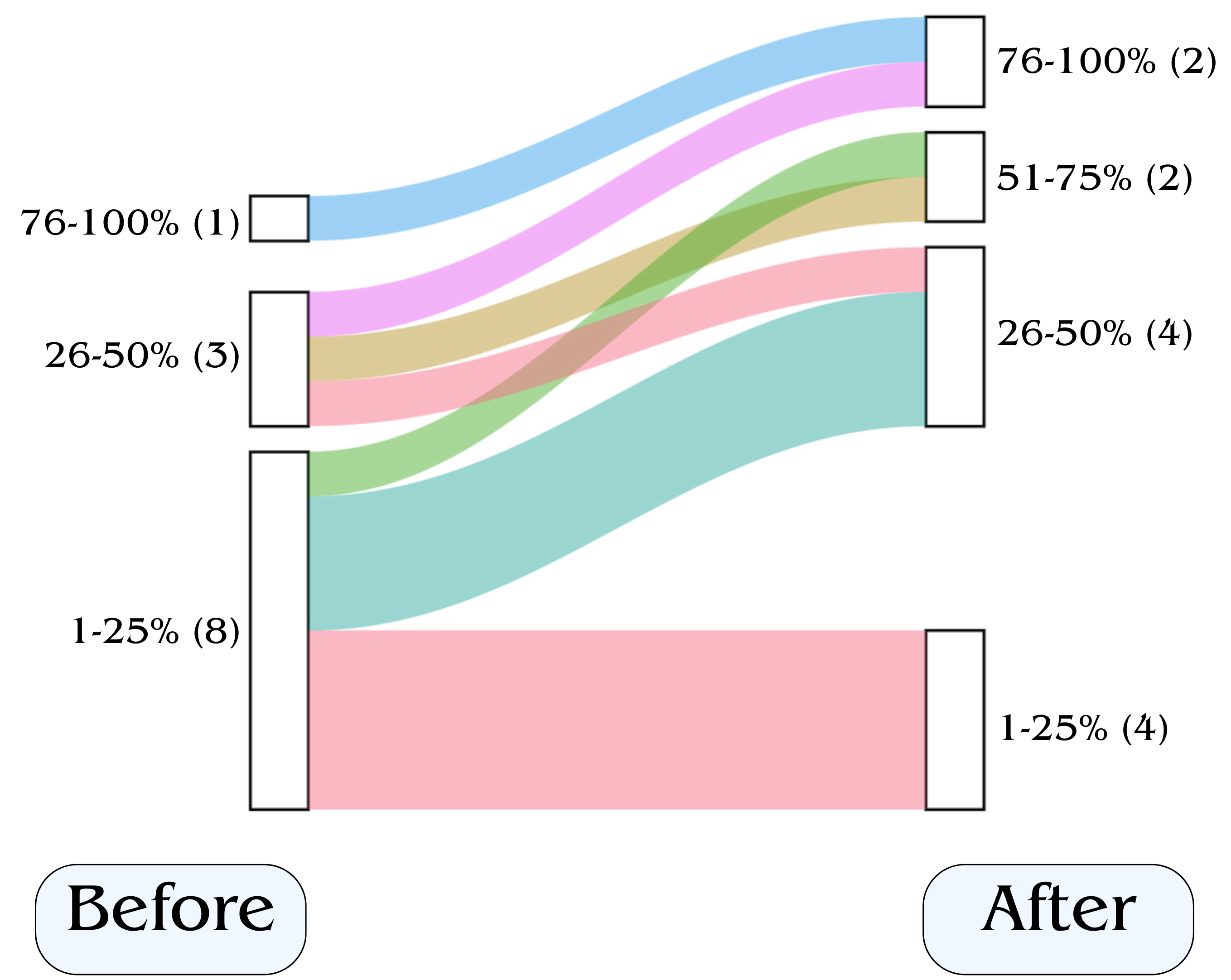}
    \caption{\sysname{}: Change in self-reported frequency of in-situ reflexivity. The Sankey diagram visualizes how participants perceived their practice of pausing to write a rationale for a code choice, comparing their recalled frequency from their typical workflows (``Before'') with their reported frequency during the study using \sysname{} (``After'').}
    \label{fig:study-reflexive-compare}
\end{figure}

Integrating structured, lightweight reflection directly into the coding interface via the \emph{ReflexiveLens} prompted a perceived shift in participants' analytical practices, from high-level, delayed habits toward more granular, in-situ engagement with their data. To investigate this perceived shift, we asked participants to self-report the frequency of their in-situ reflection for both their typical workflows and their work during the study. While we acknowledge that comparing recalled behavior to in-the-moment experience is not a controlled measure, it provides insight into the participants' own sense of their changing practice. As illustrated in Figure \ref{fig:study-reflexive-compare}, this self-reported data shows a notable trend toward more frequent reflection when using \sysname{}.

This finding aligns with our other qualitative data. During the study, participants' interaction with \textit{ReflexiveLens} was largely smooth and frictionless after their initial encounter. A majority of participants also reported that, compared to their typical workflow, the tool prompted them to pause and articulate their rationales more often. Participants commended the system's novelty in providing a dedicated and contextualized space for reflexivity, although their feedback also highlighted challenges regarding workflow and the scope of reflection, which we outline below in Section \ref{sec:result-reflexive-4}.

\subsubsection{Providing a New, Integrated Space for Situated Reflection}
\label{sec:result-reflexive-2}

A primary benefit of \textit{ReflexiveLens} was its ability to address the critical challenge of disconnected and unorganized reflective workflows. Participants frequently compared the tool's integrated nature favorably against their current manual methods. $P_{11}$, for example, described her typical process of keeping a ``side document'' for reflections, which she found \textit{``really difficult to be going back and forth, like, trying to link excerpts''} and made her prone to losing context over time. 

For other participants, the tool provided a dedicated space for a practice they valued but struggled to incorporate into their existing toolchains. As $P_4$ noted, the inline prompts were particularly useful as \textit{``otherwise, I wouldn't really have a place to do it if I'm not using specialized tools''} ($P_4$). Despite the value of having an integrated space, the interaction design sometimes created friction. $P_3$ felt the multi-step workflow of clicking through several screens to log a reflection \textit{``diverted my attention to the interface, rather than what I was thinking of,''} and suggested that a more streamlined interaction could better support rapid thought capture ($P_3$).

\subsubsection{Shifting Reflection from a Broad, Delayed Act to a Deliberate Practice}
\label{sec:result-reflexive-3}
The most novel aspect of \textit{ReflexiveLens} was its ability to facilitate reflection at the snippet level, a significant departure from most participants' existing habits of reflecting broadly on an entire interview \textit{``at the end of a coding session''} ($P_7$). For instance, we observed $P_7$ use the \textit{ReflexiveLens} prompt immediately after applying an `emotional response' code to a passage describing an ``angry'' and ``apologetic'' resident. In her reflection, she articulated a direct link between her coding choice and her positionality as a woman. 
This behavior illustrated how the tool transformed a typically delayed, high-level task into an immediate, code-specific, and deeply personal analytical act.

The tool's ability to support reflection on specific codes as they were applied was considered a valuable improvement. This granular, in-the-moment approach was not just a siloed activity; it also scaffolded collaborative sensemaking. $P_{12}$ recalled how seeing their teammate’s ($P_{11}$) reflexive notes prompted them to \textit{``think very hard''} about their reasoning, thereby encouraging deeper engagement with both the data and their own interpretive process ($P_{12}$).

These experiences are substantiated by our survey data. A large majority of participants (10 of 12) agreed that the prompts in \sysname{} encouraged them to articulate their code rationale. Crucially, this deeper reflection was well-integrated into the analytical process, nearly all participants (10 of 12) reported a neutral to positive experience with the in-flow prompts, indicating that this new way of reflecting did not interrupt their analytical workflow.

\subsubsection{Preferences in Reflective Scope: A Desire for Higher-Level, Networked Memos}
\label{sec:result-reflexive-4}
While the ability to reflect on a granular level was highly valued, the tool's per-passage focus did not perfectly align with the workflow of all participants. Some participants ($P_1$, $P_5$, $P_6$, $P_9$) described that their most critical reflections occur at a higher level of abstraction. $P_6$ found it difficult to be meaningfully reflexive on isolated passages, noting, \textit{``I think of it more in terms of how I apply multiple codes over time.''}  $P_9$, whose memoing process typically happened at the theme-level, expressed a need for features that support relational analysis:

\begin{quote}
    \textit{``if I have two codes and I realize that there's a contradiction here, I want to make a note that connects both of those codes together... that ability to kind of network my thoughts together.''} ($P_9$)
\end{quote}

In summary, \textit{ReflexiveLens} was seen to be effective at introducing and scaffolding a more granular, in-the-moment reflective practice, which participants reported as novel and useful for preventing context collapse and promoting discussion. The overall sentiment towards the feature was positive, with over half of the participants (8 of 12) deciding they would keep the reflective prompts enabled in real projects. The reported challenges with workflow friction and the expressed desire for features that support higher-level, relational reflection are key findings that point to the diverse needs of qualitative researchers.


\subsection{Making Code Evolution Transparent Prompted Deliberate and Critically Reflective Analysis \transparentcolor{(RQ2)}}

Making code evolution transparent via the \emph{Analysis History} was a significant improvement over participants' current workflows. All participants found this feature directly supported methodological transparency and collaboration. The proactive \emph{Code Drift Alert} was more polarizing: while it prompted valuable critical reflection on code definitions for some, its interruptive nature created friction for others.

\subsubsection{The Analysis History Was Universally Embraced as a Mechanism for Transparency and Rigor}
The \emph{Analysis History} feature was met with strong enthusiasm across all participant pairs. This positive reception was confirmed in our survey, where 11 of 12 participants agreed that the feature increased the transparency of their research process and allowed them to easily see code changes over time.
The feature's primary value, as described by participants, was its ability to automate and centralize a process that is currently manual, fragmented, and tedious. Researchers described that their ad-hoc workflows involved \textit{``a Google slideshow, my Atlas.ti, and a couple Google Doc memos''} ($P_5$), or tracking changes via the ``clunky'' edit history in Google Sheets ($P_6$). The system's ability to provide a visual and integrated history was seen as a significant improvement. $P_{12}$, whose team currently tracked changes manually in an Excel sheet, saw it as a powerful automation of a necessary but laborious task.

Participants identified two high-impact use cases for this newfound transparency. First, researchers directly connected the feature to the task of writing up their methodology. $P_3$ explained that the visual graph is a \textit{``very useful way to think back and see how different codes merged or split up,''} which would help in reporting the analytical process. $P_4$ described the feature as an audit trail for their work:

\begin{quote}
    \textit{``I think it's definitely going to be helpful, for transparency's sake. It's like you have a blockchain or something equivalent, so you have like, a universally agreed upon, history of changes.''} ($P_4$)
\end{quote}

Second, the history was also seen as a crucial tool for teamwork. $P_9$ described it as ``insanely useful'' for asynchronous teams and for getting a \textit{``new coder coming into a project up to speed.''} $P_3$ added that it would be \textit{``a good way to mediate those discussions''} with supervisors and collaborators. $P_{12}$ emphasized its role in team accountability, noting, \textit{``we can look back at any point in time and see what was the decision.''}

\subsubsection{The Code Drift Alert Sparked Critical Reflection, Highlighting the Value of Proactive AI Partnership}
The \emph{Code Drift Alert} introduced a novel capability for proactively monitoring analytical consistency.  While survey responses indicated that the feature helped many articulate their code's boundary (7 of 12), the qualitative feedback was more divided on its timing and interruptive nature. 

Our technical review of the study logs verified that 100\% of the displayed drift alerts were relevant, in that they flagged demonstrable linguistic or semantic variations between the original definition and the new usage. For several participants, the alert was a highly effective intervention. For instance, upon receiving an alert, $P_3$ was observed pausing his work. He first noted that the AI's observation was ``interesting,'' and after a moment of consideration, concluded, ``Actually, I agree.'' He then directly incorporated the AI's suggested wording into his code definition, later reflecting that the intervention gave him a connection \textit{``that I had not thought of before.''} $P_{11}$ also saw it as a critical tool for maintaining rigor over long projects, especially to counteract the effects of researcher fatigue:
\begin{quote}
    \textit{``especially after a certain level of fatigue, you might just start, kind of, sloppily applying a certain code. And, having it kind of flag that... can also make you self-aware... and also keep you, sort of, more precise in how you're selecting codes.''} ($P_{11}$)
\end{quote}

In contrast, other participants found the proactive, interruptive nature of the alert to be a point of friction during their focused analytical work. $P_9$ had a strong negative reaction to the unsolicited pop-up, asking it to \textit{``stop interrupting me in the middle of [coding].''} $P_2$ voiced a similar concern about the timing, suggesting the \emph{Code Drift Alert}'s value might be greater if it functioned as a review tool that could be invoked at a different stage:

\begin{quote}
    \textit{``It would have been more useful if it showed up at the end of a document or have it at the end of, like, a coding session, just to review what you've coded.''} ($P_2$)
\end{quote}
Despite the mixed feedback on its timing, participants demonstrated sustained engagement after the alert was shown. This was visible even when a participant was taken aback. For instance, after an alert ``\textit{took [her] by surprise,}'' $P_2$ was observed immediately pausing to refine her code's definition. $P_1$’s actions further illustrated this pattern of deliberate engagement. She described accepting the alert’s suggestion to create a new code on one occasion, while on another, she confidently overrode it.


\subsection{Scaffolding Collaborative Interpretation by Focusing and Framing Disagreement \collaborationcolor{(RQ3)}}

To understand how our system scaffolds collaborative interpretation (RQ3), we analyzed how research pairs engaged with features designed to surface and contextualize interpretive differences. The findings showed that the \emph{Discussion Focus} was seen as a practical enhancement to existing workflows. The \emph{Positionality-Aware Discussion Prompt}, although a more context-dependent feature, was recognized for its unique potential to foster deeper, methodologically-grounded reflexive dialogue.

\subsubsection{Discussion Focus Enhanced Collaborative Workflows by Automating Disagreement Discovery.}
Participants immediately identified the \emph{Discussion Focus} features as a practical and powerful tool for structuring collaboration. This was one of the most highly rated features in our survey, with 11 of 12 participants agreeing that it helped them notice meaningful disagreements and made it significantly easier to start or engage in discussions with collaborators. This was evident in pairs' immediate adoption of the feature to structure their conversation. For instance, at the start of their discussion, $P_2$ announced, \textit{``I just toggled on [the show diff view toggle]... so we're just gonna focus on that,''} while $P_{11}$ began by acting on the discussion status icon's information: \textit{``I'm gonna start with the one that says, needs discussion.''}

Participants described the feature as ``fundamental'' to the coding process ($P_{12}$) and \textit{``very useful, because it kind of just brings out those conflicts, very quickly''} ($P_3$). $P_5$ contrasted the tool's efficiency with her current ad-hoc method of keeping a separate list in a Google Doc and then having to \textit{``re-find the code, whenever we have a meeting to go over code disagreements''} ($P_5$). The most impactful finding was that by lowering this barrier, the tool enabled entirely new, higher-quality workflows that participants felt were previously out of reach due to time constraints:

\begin{quote}
    \textit{``[An ideal productive discussion] usually doesn't happen, at least for me it hasn't happened, because we were rushing into reaching a deadline, this flagging of discussion points could initiate [or] could give you the opportunity that, yeah, let's discuss it.''} ($P_{10}$)
\end{quote}

This sentiment was echoed by $P_9$, who confirmed that the feature would \textit{``take you from not discussing these things to discussing them, which will increase the quality of the analysis''} ($P_9$). By directly addressing the logistical challenges of manual comparison, the feature was seen as a new and useful tool for making collaboration more efficient and rigorous.

\subsubsection{Positionality-Aware Prompt: A Novel Framing for Disagreement with Context-Dependent Utility.}

The AI-powered \emph{Positionality-Aware Discussion Prompt} elicited a more mixed reaction. Participants recognized its unique value in fostering methodologically-grounded reflection, even as they noted practical limitations in its current form.

Our technical review affirmed that all generated prompts were relevant, explicitly referencing the conflicting codes and the users' self-authored background and positionality context. Likewise, participants found the system generally provided a safe and comfortable environment to express disagreement. This potential was recognized immediately by some. Upon first seeing the prompt during the study, $P_7$ offered a spontaneous, in-the-moment analysis of its value for supporting an interpretivist paradigm:

\begin{quote}
    \textit{``[The Discussion Prompt] is actually really nice... it does show that there's a disagreement, but it shows it in a way not that, oh, this is the actual ground truth with interpretivism, we accept that there can be multiple truths or multiple realities, I like that this is shaping it as this person has this positionality, and that's why you might have different codes.''} ($P_7$)
\end{quote}

This sentiment was supported by $P_{11}$, who found the AI prompt demonstrated a ``strong understanding'' of the disagreement and provided a \textit{``good prompt for that kind of discussion, because after a long amount of coding, everything can sort of start blurring together''} ($P_{11}$).

Participants also pointed out the limitations of the Discussion Prompt, primarily its shortcomings in facilitating synchronous collaboration. We observed pairs bypassing the AI in favor of direct conversation, as they found it to be more convenient. Participants critiqued the AI's output as \textit{``too long and too generic''} ($P_2$) and noted that it failed to meaningfully incorporate their positionality ($P_4$). The prompt's value became clearer, however, when participants considered asynchronous work. \textit{``I think it would be more useful than… when we're chatting now,''} stated $P_1$, highlighting that the feature's utility is highly dependent on the collaborative context.


\subsection{Overall User Experience}
Beyond the findings for specific features, our study revealed a strong positive reception of \sysname{} as a complete system, positioning it as a significant improvement over participants' existing qualitative analysis workflows.  This was supported by the unanimous agreement from participants on the system's core principles: every single participant (12/12) reported feeling ``in control of analytic decisions'' and agreed the system ``supported an iterative research process.''

Participants frequently attributed this holistic improvement to the tool's integrated nature, which removed the cognitive friction of their current, fragmented workflows. $P_7$, for instance, contrasted \sysname{} favorably with their ``document-heavy'' Google Docs method. The tool's ability to facilitate collaboration was a standout benefit, addressing a critical gap in the current software landscape. As $P_5$ explained:
\begin{quote}
    \textit{``I would use it. There is no good open, like, free lightweight online coding tool that exists. Like, you can use [other tools], but then you can't resolve disagreements easily, between multiple coders.''} ($P_5$)
\end{quote}
Even when compared to specialized software like NVivo, participants praised \sysname{} for having a less ``clunky'' ($P_{12}$) and more ``elegant'' ($P_9$) interface, making the core coding process ``a lot more pleasant'' ($P_6$).

Of course, the experience was not without limitations. Some participants noted the study's time constraints created a feeling of pressure, preventing them from the deeper, more time-intensive engagement the tool affords ($P_9$). The most common requests for improvement were visual, including color-coded highlights and a hierarchical codebook to better organize themes ($P_7, P_{11}, P_{12}$).
We report the detailed descriptive statistics in Appendix \ref{appendix:user-study-quant-stats}.

Taken together, our findings demonstrate that by thoughtfully integrating support for reflexivity, transparency, and collaboration, computational tools can move beyond simple code management to become scaffolds for a more deliberate and reflexive analytical process.
\section{Discussion: Designing for Deliberation in Human-AI Systems to Support Human Collaboration}
As AI models increasingly automate complex cognitive tasks, the challenge for HCI is shifting from designing efficient human-AI pipelines to a more intentional goal: \textbf{\emph{designing for deliberation}}. This approach, which builds on calls to treat ambiguity as a design resource \cite{gaver2003ambiguity} and to create technologies that support reflection \cite{baumer2015reflective}, prioritizes systems that intentionally support the deep, and often contentious, processes of human interpretation \cite{ma2025towards}. It stands in contrast to an uncritical focus on automation, which risks diminishing creative originality and suppressing critical thought \cite{kumar2025human, 10.1145/3706598.3713778}.

In this paper, we operationalize the concept of \textit{design for deliberation} to address the specific methodological challenges of collaborative reflexive analysis, a practice where interpretive depth and methodological rigor are paramount. Through the design and evaluation of \sysname{}, we demonstrate how computational tools can scaffold this deliberate practice by surfacing latent tensions and maintaining interpretive context across discontinuous workflows. Our findings suggest that this approach offers a viable path for preserving methodological rigor in the age of automation, specifically by prioritizing researcher agency, transforming implicit context into explicit history to support task resumption, and navigating the epistemic risks of operationalizing human positionality in computational systems.

\subsection{Augmenting Agency: Shifting from Automation to Deliberation}
Prevalent paradigms in Human-AI interaction often prioritize the reduction of cognitive load, positioning AI as a tool to automate tasks and streamline efficiency. However, in the context of qualitative inquiry, we argue that this focus on ``computational convenience'' is misplaced. Our work demonstrates the value of an alternative interaction paradigm: using AI's analytical capabilities not to replace human judgment, but to provoke critical self-awareness during complex knowledge work. \sysname{} operationalizes this approach by functioning as an analytic scaffold rather than an oracle, surfacing potential inconsistencies to stimulate, rather than resolve, researcher deliberation.

This approach addresses emerging concerns regarding the epistemic risks of over-automating cognitive work. Recent research indicates that reliance on generative models can lead to the ``homogenization'' of creative outputs \cite{kumar2025human}, while AI-generated explanations may paradoxically suppress critical thinking and promote overreliance \cite{10.1145/3706598.3713778, kim2025fostering}. In qualitative analysis, where the scientific value is derived from the depth of the researcher's interpretive effort, such automation is particularly counterproductive \cite{dourish2014reading, jowsey2025we}. \sysname{} mitigates these risks by casting the AI as a non-prescriptive observer. For instance, the Code Drift Alert does not ``correct'' the researcher but rather flags potential concept drift to force a deliberate choice. Participants reported that this friction made them ``more precise'' and ``self-aware''($P_{11}$), illustrating how AI can be designed to deepen, rather than bypass, intellectual rigor.

This design pattern, which employs AI to encourage deeper immersion of context, surface latent tensions, allow reasoned deliberation, and even demand human-led critical inquiry, has broad implications for other domains of interpretive and creative work:
\textit{For Scholarly and Creative Writing} $\Rightarrow$  An AI system could act as a developmental editor by tracking the evolution of arguments or character arcs in a long-form document. It would highlight potential inconsistencies or tonal shifts for the author's review, preserving their narrative authority while augmenting their capacity for self-critique, designed to deepen, rather than bypass, intellectual rigor. \textit{For Design and Engineering} $\Rightarrow$ An AI could juxtapose a designer's specifications against their stated goals or established design patterns. It could then surface tensions, for example, a conflict between a requirement for ``ease of use'' and an architectural choice that adds complexity. This prompts the designer to explicitly rationalize trade-offs early in the implementation process.

\subsection{Designing for Analytical Provenance to Support Resumption and Longer-term Usage}

To ensure the long-term usability of qualitative tools in iterative knowledge work, systems must address the temporal reality of research projects that extend over weeks or months, rather than single sessions \cite{braun2006using}. However, a primary challenge in longitudinal qualitative analysis is the cognitive cost of \textit{resumption}, characterized by the difficulty of recovering interpretive details when researchers return to analysis after periods of downtime \cite{lipford2010helping}. 

To address this, we draw on prior visualization research that distinguishes between \textit{interaction provenance} (logs of user events) and \textit{rationale provenance} (the reasoning behind decisions) \cite{ragan2015characterizing, xu2020survey}. While interaction logs support reproducibility, they often fail to support the cognitive task of \textit{resumption}, as raw event streams lack the semantic meaning necessary for a researcher to recover their state of mind weeks later \cite{lipford2010helping, rule2018aiding}.

\sysname{} is designed to bridge this gap by prioritizing a meaningful, auditable narrative of conceptual development over an exhaustive interaction log. We operationalize this via \textit{data-level provenance} -- which visualizes the structural evolution of analytic actions (e.g., splits, merges, renames) \cite{viegas2004studying}, augmented by \textit{lightweight rationale provenance} -- the brief reflections and justifications linked to individual snippets. Mirroring narrative scaffolding work that suggests insight-linked provenance helps analysts act on reflections~\cite{huang26narrative}, our findings (RQ2) indicate that combining data-level provenance with lightweight rationales effectively balance utility and manageability. By transforming the codebook from a static list into a living history, the system shows potential as an external memory scaffold and is ``a very useful way to think back.'' This allows researchers to combat the ``drift'' and fatigue inherent in longitudinal work \cite{o2020intercoder}. While longitudinal deployment is necessary to confirm long-term adoption, this immediate utility suggests that provenance views within \sysname{} can facilitate re-immersion and transparent reporting, without the overhead of manual journaling.

\subsection{Navigating Epistemic Risks in Operationalizing Positionality}
Expertise in knowledge work is informed by a rich context of prior experiences, theoretical commitments, and rationales. While most systems capture the outputs of an action (e.g., codes applied, decisions made), the interpretive context shaping those actions often remains implicit and computationally inaccessible. \sysname{} operationalizes this context by treating positionality as active data to anchor discussion prompts and surfaces the analyst’s internal stance \cite{dourish2014reading}.

However, translating complex human positionality into computational prompts introduces consequential epistemic risks. As highlighted by the inherent brevity of user profiles, there is a tension between making positionality computationally tractable and preserving the nuance of a researcher's entire lived experience. As we move toward systems that model user perspectives, there is a danger of algorithmic essentialism, which reduces a researcher to a set of flat characteristics or stereotypes. Large language models, which rely on probabilistic correlations, may inadvertently generate reductive or essentializing attributions (e.g., suggesting a researcher's interpretation is solely defined by a single demographic trait rather than their holistic expertise). This further risks reifying stereotypes or entrenching assumptions of how people with certain identities or perspectives interpret data,  undermining the researcher’s authority and contextual information.

To mitigate this, we argue that systems must be designed to \textit{operationalize positionality as inspiration, not diagnosis}. In \sysname{}, the prompts are framed as tentative conversation starters rather than declarative truths. As mentioned by our study participant, this interaction design shifts the epistemic burden back to the human; the system provides a heuristic spark to disrupt pre-existing perspectives, but the researcher retains the agency to interact, reject, or recontextualize that connection.

Operationalizing and surfacing user perspectives in this way enables systems to facilitate more subtle interactions in high-stakes tasks, such as academic peer review. Despite showing signs of enhancing peer review quality \cite{thakkar2025can}, current research primarily focuses on efficiency \cite{chen2025envisioning, hossain2024llms}. But a system that models the reviewer's perspective could address the challenge of interpretive diversity. For a meta-reviewer, such a system could contextualize conflicting assessments by highlighting how reviewers' differing theoretical lenses shaped their critiques. However, our findings suggest a critical constraint for such future tools: to avoid reductive profiling, they must remain tentative and non-prescriptive, surfacing potential associations between diverse backgrounds, acting only to stimulate human deliberation.
\section{Limitations}
Our study and system have several limitations that bound our claims. First, the evaluation involved 12 experienced qualitative researchers working in pairs within a single, time-limited session on a synthetic dataset. This setting let us observe rich, situated use of \sysname{} and provided interaction-level insights into reflexivity features, but it did not capture the non-linear, longitudinal dynamics of real projects that unfold over months, nor the influence of institutional constraints, shifting research questions, or evolving team relationships. As such, our findings speak to initial use and perceived affordances rather than long-term change in practice.

Second, we did not study larger teams or strong power asymmetries. While \sysname{} technically supports multi-party collaboration, we do not yet know how features like Discussion Focus or positionality-aware prompts play out when three or more interpretive lenses, organizational hierarchies, or disciplinary differences are in tension.

Finally, our operationalization of positionality is necessarily partial. \sysname{} operationalizes positionality as brief, self-authored profiles that are summarized into prompts. This makes positionality computationally tractable but risks flattening complex, intersectional, and evolving positions into relatively stable descriptors. We did not systematically examine potential harms or discomfort this might create for marginalized researchers or in sensitive domains, so we interpret our results as evidence of both promise and risk rather than as validation of positionality-aware prompting as a general best practice.

\section{Future Work}
Several directions follow from this work. First, we plan to conduct longitudinal deployments of \sysname{} with teams using the system throughout the lifecycle of research projects. Such studies can examine how reflexive streams, provenance views, and positionality-focused collaboration are adapted, incorporated, or abandoned over time, and whether they ultimately reshape how findings are constructed and reported.

Second, we aim to co-design alternative ways of representing positionality in collaborative tools. Rather than a single static profile per researcher, teams might work with evolving or context-specific positionality traces, shared team-level statements, or prompts that explicitly invite critique of the system’s own attributions. We are particularly interested in partnering with researchers from marginalized communities and with teams operating under pronounced power asymmetries to understand when positionality-aware prompts are helpful, when they are risky, and what forms of control, consent, and redress are needed.

Third, we see opportunities to extend \sysname{} along infrastructural lines, including support for locally hosted language models and configuration options that let teams selectively enable or disable LLM-powered features. More broadly, we hope to develop design guidelines and evaluative frameworks for ``designing for deliberation'' in interpretive human–AI systems.

\section{Conclusion}
This paper introduces \sysname{}, a collaborative workspace that deeply embeds researcher reflexivity, code evolution, and principled disagreement into the analytical workflow. Through a paired-analyst study with 12 qualitative researchers, we showed how integrating in-situ reflexive streams, analytical provenance, and disagreement-centered collaboration can encourage a more granular, deliberate, and methodologically transparent practice than is supported by many current toolsets and workflows. Our findings exemplify how AI can scaffold reflexive and dialogic practices. At the same time, the study surfaces important risks and limits of operationalizing positionality in tools, while effective prompts can anchor productive dialogue, generic or misaligned framings risk flattening complex identities and underplaying reflexive work. We offer \sysname{} as a design probe towards the challenges of positionality in collaborative qualitative analysis. By intentionally designing for deliberation rather than efficiency, our work contributes a principled approach to building human-AI systems for interpretive knowledge work. The insights from \sysname{} can guide the future tools for thought that honor and augment, rather than automate, the deep, personal processes of human sensemaking and knowledge-seeking.

\begin{acks}
We extend a heartfelt thanks to our study participants, without whom this work would not have been possible.

We acknowledge and thank the support of the Natural Sciences and Engineering Research Council of Canada (NSERC), [funding reference number RGPIN-2024-04348 and RGPIN-2024-06005]. This work was also supported by the National Science Foundation under Award No. 2402428.
\end{acks}

\bibliographystyle{ACM-Reference-Format}
\bibliography{reference}

\appendix

\onecolumn
\section{Appendix}
\subsection{Formative Study (Prolific Survey)}
\label{appendix:formative-survey-details}
Prolific participant pool filtering criteria: Highest education level completed: Graduate degree (MA/MSc/MPhil/other), Doctorate degree (PhD/other).

Prolific screening questions verified that participants had
\begin{enumerate}
    \item completed at least one thematic analysis project
    \item familiarity with thematic analysis methodology
    \item a minimum of 2 years of qualitative analysis experience
\end{enumerate}

The survey was structured in four sections:
\begin{enumerate}
    \item background and experience (8 questions)
    \item reflexive practices during coding (12 questions using 5-point Likert scales)
    \item code evolution tracking strategies (7 questions combining multiple choice and open-ended responses)
    \item collaborative interpretation challenges (9 questions, including ranking and free-text).
\end{enumerate}

Survey response statistics: Approved (55), Returned (13, including 2 survey participants who have completed the survey but manually returned due to exceptionally low effort showed), Screened out (12), Timed-out (1). Median survey completion time 00:27:39.

\newpage
\subsection{LLM System prompt used in \sysname{}}
\label{appendix:llm-prompt}
\lstset{
    breaklines=true,                 
    basicstyle=\ttfamily\footnotesize,        
    backgroundcolor=\color{gray!10},
    breakatwhitespace=false,         
    showstringspaces=false,          
    frame=single,                      
    breakindent=0pt,                 
    language=,                       
    escapeinside={(*@}{@*)},         
}

\noindent\textbf{Code Drift Alert}
\begin{lstlisting}
You are a sharp-eyed assistant for reflexive qualitative researchers. Your task is to spot 'conceptual drift' and flag it with a brief, scannable alert.

Conceptual drift is when a new passage stretches or shifts a code's meaning beyond its current definition and applied examples.

Your response must be extremely concise. The goal is to provide a quick "heads-up" to the researcher, not a detailed analysis. Use plain language and get straight to the point. When drift is detected, your explanation should clearly and simply contrast the original concept with the new one.

-----

Current definition: ${codeDefinition}

Existing passages coded with ${codeName}:
${examplesText}

NEW PASSAGE (selected portion): ${newPassage}, which is situated within the bigger context ${context}.

Does this new passage represent conceptual drift?

Respond in JSON format.
- "drift_detected": boolean
- "explanation": A single, concise sentence contrasting the original focus with the new usage. If no drift, state that the usage is consistent.
- "suggested_definition": A revised definition if drift is detected, otherwise null.

response_format:
{
  "type":"json_schema",
  "json_schema":{
    "name":"conceptual_drift",
    "schema":{
      "type":"object",
      "properties":{
        "drift_detected":{
          "type":"boolean",
          "description":"Indicates if conceptual drift was detected"
        },
        "explanation":{
          "type":"string",
          "description":"Brief explanation of the drift or why no drift was detected"
        },
        "suggested_definition":{
          "type":"string",
          "description":"Revised definition if drift detected, otherwise null"
        }
      },
      "required":[
        "drift_detected",
        "explanation"
      ],
      "additionalProperties":false
    }
  }
}
\end{lstlisting}

\newpage
\noindent\textbf{Positionality-Aware Discussion Prompt}
\begin{lstlisting}
You are an expert facilitator for reflexive qualitative research teams. Your task is to generate brief, engaging "Conversation Starters" when researchers code the same text differently, possibly due to different interpretations and backgrounds.

Your output must:
1.  **Be extremely concise.** The goal is to spark a live conversation, not to be read at length.
2.  **Ask a direct question.** Do not provide a pre-packaged analysis. Your job is to help the researchers discover the insight themselves.
3.  **Frame the difference in interpretation and background as a valuable tension**, not a problem to be solved.
4.  **Use plain, conversational language.** Avoid academic jargon.
5.  **Directly address the researchers** by their first names to make it personal.
6.  Generate a **short, memorable title** that captures the essence of the coding difference.

-----

Generate a "Conversation Starter" for qualitative researchers who coded the same text differently.

FULL CONTEXT: ${context}

CODED TEXT (selected portion): ${codedText}
${documentTitle}

RESEARCHERS AND CODES:
${researcherDescriptions}

CODE DEFINITIONS:
${codeDefinitionsText}

Based on the information above, generate a title and a prompt. The prompt should be a single, direct question.
- The **Title** should be a short phrase capturing the tension (e.g., "Process vs. Price").
- The **Prompt** should be a single, curious question that encourages the researchers to reflect on their perspectives. Address them by their first names.


response_format: 
{
  "type":"json_schema",
  "json_schema":{
    "name":"discussion_prompt",
    "schema":{
      "type":"object",
      "properties":{
        "prompt":{
          "type":"string",
          "description":"The generated discussion prompt"
        },
        "title":{
          "type":"string",
          "description":"A brief title for the insight opportunity"
        }
      },
      "required":[
        "prompt",
        "title"
      ],
      "additionalProperties":false
    }
  }
}
\end{lstlisting}

\newpage
\noindent\textbf{Reflexive Stream Structured Summary}
\begin{lstlisting}
You are a reflexive research coach. Your purpose is to act as a mirror, helping researchers see the patterns in their own thinking based on their reflexive notes.

Your task is not to judge their analysis, but to synthesize their entries into concise observations and thought-provoking questions. The goal is to deepen their self-awareness as an instrument of their research as a team.

Analyze the notes across these dimensions:
1.  **Linguistic Patterns:** How they use language to justify interpretations.
2.  **Positionality Narrative:** The story they tell about their own background and biases.
3.  **Alternative Thinking:** The ways they do (or don't) challenge their own conclusions.

The tone should be supportive, curious, and collaborative. Address the team instead of individuals and frame your outputs as "Reflective Starting Points.

-----

Please analyze the following reflexive responses from a reflexive qualitative research team:

JUSTIFICATION RESPONSES (What specific language led to coding decisions):
${justificationResponses.map(...)}

POSITIONALITY RESPONSES (Personal/professional experiences influencing interpretation):
${positionalityResponses.map(...)}

ALTERNATIVE FRAMING RESPONSES (Different possible interpretations):
${alternativeResponses.map(...)}

OTHER NOTES:
${notes.map(...)}

Please generate a set of "Reflective Summary". For each category, provide a concise one-sentence finding and a follow-up reflective question to prompt deeper thought. Address the researchers directly as 'you' or 'your team'.


response_format:
{
  "type":"json_schema",
  "json_schema":{
    "name":"reflexive_summary",
    "schema":{
      "type":"object",
      "properties":{
        "linguisticPatterns":{
          "type":"string",
          "description":"A one-sentence observation about language patterns for this particular code."
        },
        "positionalityNarrative":{
          "type":"string",
          "description":"A one-sentence synthesis of the team's positionality on this particular code."
        },
        "alternativeThinkingPatterns":{
          "type":"string",
          "description":"A one-sentence observation on how alternatives are generated."
        },
        "notes":{
          "type":"string",
          "description":"A one-sentence summary of additional notes and context provided by researchers."
        }
      },
      "required":[
        "linguisticPatterns",
        "positionalityNarrative",
        "alternativeThinkingPatterns",
        "notes"
      ],
      "additionalProperties":false
    }
  }
}
\end{lstlisting}

\newpage
\noindent\textbf{Positionality Keywords}
\begin{lstlisting}
You are an assistant that produces concise, semicolon-separated keyword summaries of a researcher's background and positionality for collaborative qualitative analysis tools.

-----

Summarize the following research background filled out by researcher into a compact list of 1-12 keywords/short phrases separated by semicolons + space (; ). 
Only return the keywords line, no extra words, labels, or quotes. Keep each keyword under 3-5 words. Prioritize items like: discipline, methods, theoretical lenses, domain expertise, populations/contexts, positionality/identity factors, likely biases, and analytic style.

Number of keywords should be proportional to the length and complexity of the research background user provided, less provided context should result in fewer than usual keywords.

${userName}
Question: Brief History of Qualitative Data Analysis?
User data: ${parsed.qualitativeHistory}

Question: How Background May Affect Interpretation?
User data: ${parsed.backgroundExperience}

Question: Initial View of the Data?
User data: ${parsed.initialDataView}


response_format:
{
  "type":"json_schema",
  "json_schema":{
    "name":"research_background_keywords",
    "schema":{
      "type":"object",
      "properties":{
        "keywords":{
          "type":"string",
          "description":"Semicolon-separated keywords that summarize the research background."
        }
      },
      "required":[
        "keywords"
      ],
      "additionalProperties":false
    }
  }
}
\end{lstlisting}

\subsection{System prompt used to generate user study synthetic dataset}
\label{appendix:llm-prompt-user-study}
\begin{lstlisting}
<insert PDF of paper "Contestable Camera Cars: A Speculative Design Exploration of Public AI That Is Open and Responsive to Dispute">

I am now conducting an interview study, and asking participants to perform a simplified Reflexive Thematic Analysis. I need to generate a dataset because there's no available data online. Can you carefully examine the paper candidate, where the author applied RTA within a very specific setting, and I want to use the paper provided as a source to create an RTA dataset using the information provided by the paper (i.e., perspective, different views, context ...).

Carefully read the provided paper, and generate "realistic" interview transcripts that fit the narrative of the paper for realistic, easy to grasp (by participants with different backgrounds, and insights do not require specialized knowledge), verbal, just like a spoken conversation, but don't make it too obvious, as we will want participants to conduct RTA

The transcript should be realistic, back and forth, and use verbal English. You want each transcript to contain a subset of themes, just like anyone would say multiple things they noticed.

Carefully think and create a narrative before returning your final answer.
\end{lstlisting}

\newpage
\subsection{User Study Participant Demographics}
\label{appendix:demographics}
\begin{table*}[ht]
\centering
\small
\begin{tabular}{llllll}
\textbf{Group \#}  & \textbf{ID} & \textbf{Current Role} & \textbf{Primary Domain}  & \textbf{Experience (Years)} & \textbf{Full RTA Comfort Level (1-7)} \\ \hline
\multirow{2}{*}{1} & P1          & Assistant Professor   & HCI                      & 5                           & 4                                \\ \cline{2-6} 
                   & P2          & PhD Student           & HCI + NLP                & 4                           & 4                                \\ \hline
\multirow{2}{*}{2} & P3          & PhD Student           & HCI                      & 5                           & 7                                \\ \cline{2-6} 
                   & P4          & PhD Student           & HCI                      & 2                           & 4                                \\ \hline
\multirow{2}{*}{3} & P5          & PhD Student           & HCI                      & 4                           & 4                                \\ \cline{2-6} 
                   & P6          & PhD Student           & HCI + Info. Vis.         & 2                           & 3                                \\ \hline
\multirow{2}{*}{4} & P7          & PhD Student           & HCI + Comp. Ed.           & 4                           & 4                                \\ \cline{2-6} 
                   & P8          & Data Analyst          & Data Allocation/Engineer & 7                           & 7                                \\ \hline
\multirow{2}{*}{5} & P9          & PhD Graduate          & HCI                      & 5                           & 4                                \\ \cline{2-6} 
                   & P10         & PhD Student           & HCI                      & 4                           & 3                                \\ \hline
\multirow{2}{*}{6} & P11         & MSc Student           & HCI                      & 2                           & 3                                \\ \cline{2-6} 
                   & P12         & Data Analyst          & Software Engineering     & 1                           & 4                                \\ \hline
\end{tabular}
\caption{User Study Participant Demographics. Participants rated their comfort level with RTA on a 7-point Likert scale, ranging from 1 (Extremely uncomfortable) to 7 (Extremely comfortable).}
\end{table*}
\begin{table*}[ht]
\centering
\small
\begin{tabular}{llll}
\textbf{Group \#}  & \textbf{ID} & \textbf{RTA Experience (simplified)}                  & \textbf{View on Topic of Analysis (simplified)}                   \\ \hline
\multirow{2}{*}{1} & P1          & Grad course; collaborative application               & HCI, User-centered, Fairness                      \\ \cline{2-4} 
                   & P2          & Four projects; varied data types                     & Immigrant, System-distrust, Cautious              \\ \hline
\multirow{2}{*}{2} & P3          & Published research                                   & AI-skeptic, Citizen-focused, Socially-critical    \\ \cline{2-4} 
                   & P4          & Coursework familiarity; partial application          & Lived experience, Pro-transparency, Balanced      \\ \hline
\multirow{2}{*}{3} & P5          & PhD course; one project                              & Urban planning, Nuanced, Anti-luddism             \\ \cline{2-4} 
                   & P6          & Applied to literature \& interviews                  & Urbanist, Anti-surveillance, Policy-first         \\ \hline
\multirow{2}{*}{4} & P7          & Applied to student perspectives                      & Interpretivist, Critical, Reflexive               \\ \cline{2-4} 
                   & P8          & Coursework \& independent training                   & Black American, Equity-focused, Data-informed     \\ \hline
\multirow{2}{*}{5} & P9          & Master's course; five studies                        & Skeptical, Accuracy-focused, Pro-supervision      \\ \cline{2-4} 
                   & P10         & PhD research; multiple methods (interviews, diaries) & International, Socio-cultural, Chatbot researcher \\ \hline
\multirow{2}{*}{6} & P11         & Initial phase reflexive coding                       & HCI/Healthcare, Immigrant-lens, Fairness-focused  \\ \cline{2-4} 
                   & P12         & Applied to dementia transcripts                      & Dementia-informed, Accessibility-focused, Pro-AI  \\ \hline
\end{tabular}
\caption{User Study Participant RTA Experience and View on Topic of Analysis \cite{camera-car}.}
\end{table*}

\newpage
\begin{figure}[H]
    \centering
    \includegraphics[width=.7\linewidth]{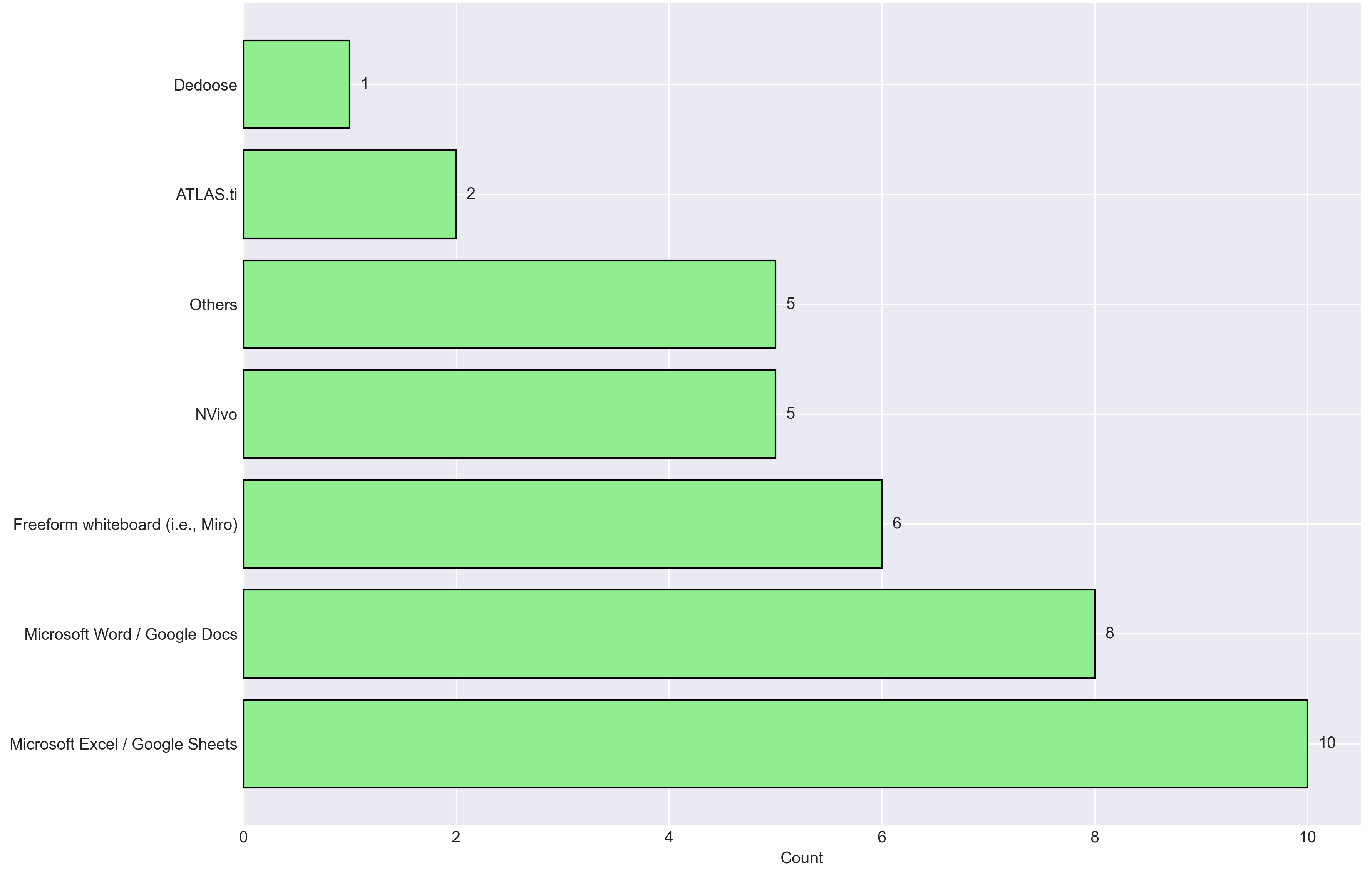}
    \caption{User Study Survey Question: What tool do you usually use for qualitative analysis?}
    \label{fig:02_tools_used}
\end{figure}
Participants who selected Others in Figure \ref{fig:02_tools_used} have shared they also use ``QualCoder'', ``printed quotes and table''
``LiquidText'', ``DiscoverText'', ``Python for extracting codes from google doc comments''.

\begin{figure}[H]
    \centering
    \includegraphics[width=.7\linewidth]{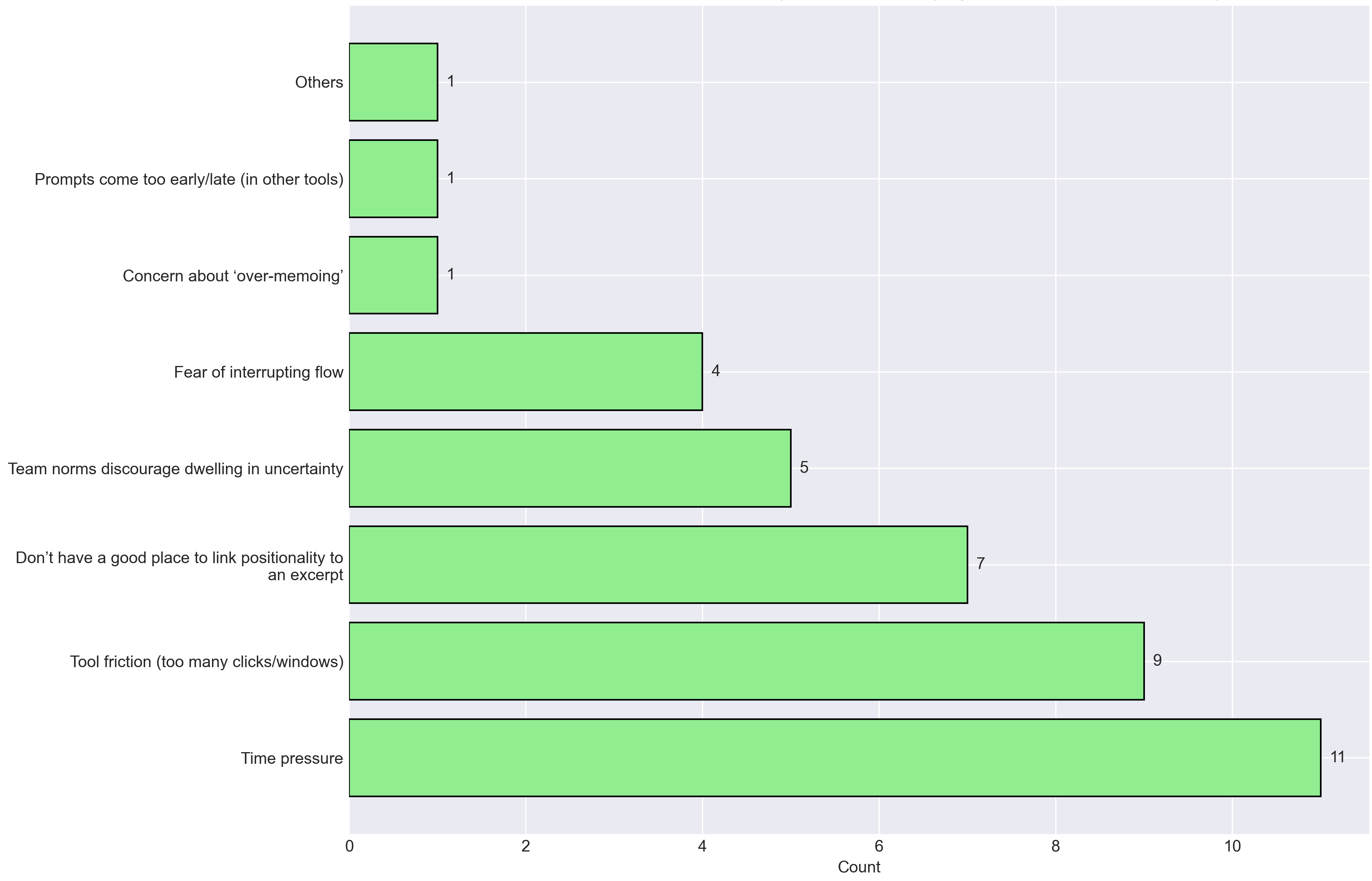}
    \caption{User Study Survey Question: What obstacles have you faced when trying to do RTA?}
    \label{fig:04_rta_obstacles}
\end{figure}
Participant who selected Others in Figure \ref{fig:04_rta_obstacles} have shared another obstacle they face is ``Not having audio and quotes in the same place''.

\begin{figure}[H]
    \centering
    \includegraphics[width=.7\linewidth]{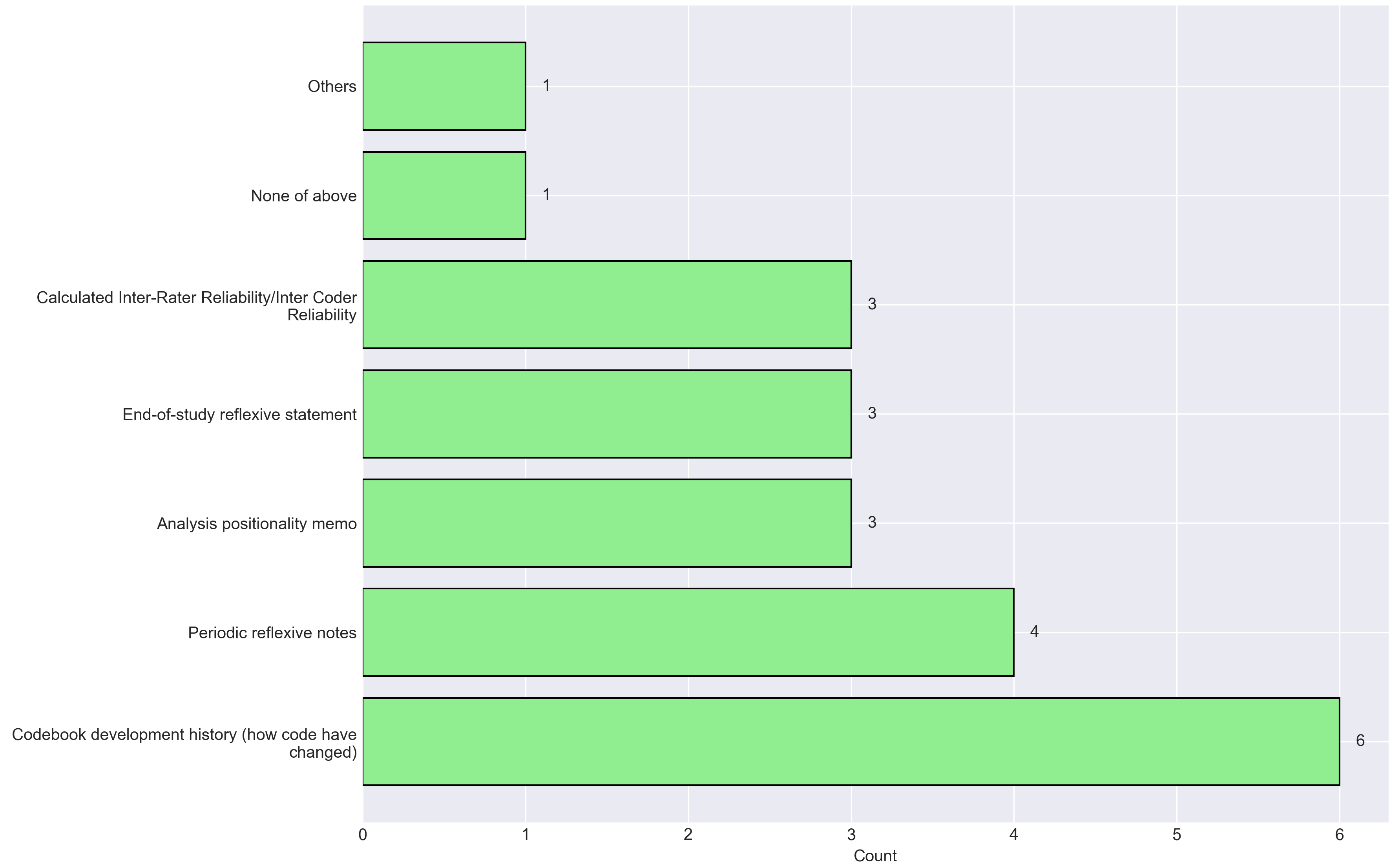}
    \caption{User Study Survey Question: In your past qualitative analysis projects, what artifacts have you created to support reflexivity and collaboration?}
    \label{fig:05_past_artifacts}
\end{figure}

\begin{figure}[H]
    \centering
    \includegraphics[width=.7\linewidth]{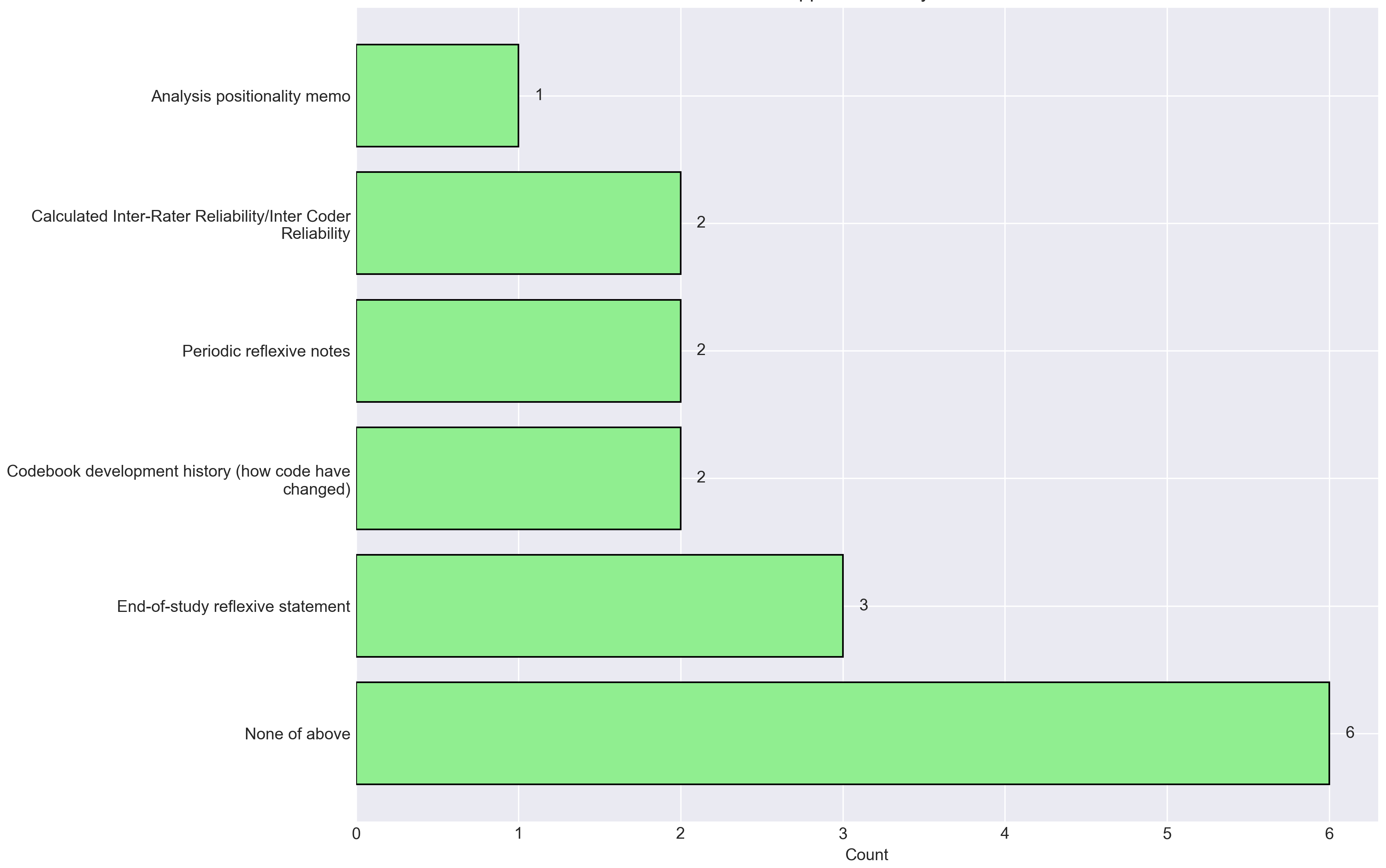}
    \caption{User Study Survey Question: In your past qualitative analysis projects, what artifacts have you \textbf{shared} with anyone outside of your research team?}
    \label{fig:06_past_artifacts_shared}
\end{figure}

\newpage
\subsection{User Study Quantitative Survey Results}
\label{appendix:user-study-quant-stats}

\begin{table*}[h]
  \caption{Aggregated Analysis of Likert Scale Questions with Categorization}
  \label{tab:user-study-qual-stats}
  \centering
  \begin{tabular}{@{} l p{9cm} c c c @{}}
    \toprule
    \textbf{Q\#} & \textbf{Question Text} & \textbf{Mean} & \textbf{Median} & \textbf{SD} \\
    \midrule

    \multicolumn{5}{@{}l}{\textbf{Supporting Reflexive Analysis (RQ1)}} \\
    \multicolumn{5}{@{}l}{\quad \textit{ReflexiveLens}} \\
    Q1 & Timely in-flow reflection & 3.67 & 4.00 & 1.30 \\
    Q2 & Articulate code rationale & 3.58 & 4.00 & 1.24 \\
    Q3 & Connect positionality to decisions & 3.42 & 3.50 & 1.44 \\
    Q4 & Consider alternative interpretations & 3.75 & 4.00 & 1.29 \\
    Q5 & Re-read data more carefully & 3.67 & 3.50 & 0.98 \\
    Q6 & Felt immersed in data & 3.50 & 3.50 & 1.17 \\
    Q7 & Would keep reflection prompts on & 3.58 & 4.00 & 1.08 \\
    \addlinespace

    \multicolumn{5}{@{}l}{\textbf{Transparent Code Evolution (RQ2)}} \\
    \multicolumn{5}{@{}l}{\quad \textit{Analysis History}} \\
    Q8 & Easily see code changes over time & 4.17 & 4.50 & 1.03 \\
    Q9 & Increased transparency of research process & 4.50 & 5.00 & 0.80 \\
    \multicolumn{5}{@{}l}{\quad \textit{Code Drift Alert}} \\
    Q10 & Surfaced potential meaningful shifts... & 3.75 & 3.50 & 1.22 \\
    Q11 & Improved the quality of our analysis & 3.50 & 3.50 & 1.38 \\
    Q12 & Helped me articulate code boundary... & 3.67 & 4.00 & 1.30 \\
    \addlinespace

    \multicolumn{5}{@{}l}{\textbf{Scaffolding Collaborative Interpretation (RQ3)}} \\
    \multicolumn{5}{@{}l}{\quad \textit{Positionality-aware Discussion Prompt}} \\
    Q13 & Reframe disagreement focus & 3.83 & 4.00 & 1.11 \\
    Q14 & Understand collaborator's positionality & 3.75 & 4.50 & 1.60 \\
    Q15 & Felt safe expressing divergent interpretation & 3.83 & 4.00 & 1.03 \\
    \multicolumn{5}{@{}l}{\quad \textit{Discussion Focus}} \\
    Q16 & Notice meaningful disagreement & 4.25 & 4.50 & 0.97 \\
    Q17 & Easy to start/engage in discussion & 4.33 & 4.00 & 0.65 \\
    \addlinespace

    \multicolumn{5}{@{}l}{\textbf{Agency}} \\
    Q18 & Felt in control of analytic decisions & 4.17 & 4.00 & 0.83 \\
    Q19 & Support iterative research process & 4.50 & 4.50 & 0.52 \\
    \addlinespace

    \multicolumn{5}{@{}l}{\textbf{Usability}} \\
    Q20 & Easy to learn & 4.00 & 4.00 & 1.04 \\
    Q21 & Easy to use & 4.25 & 4.00 & 0.75 \\
    Q22 & Follow RTA practices & 4.08 & 4.00 & 1.00 \\
    Q23 & Require less mental effort & 2.92 & 2.50 & 1.51 \\
    Q24 & Would use on real projects & 4.42 & 4.50 & 0.67 \\
    \addlinespace

    \multicolumn{5}{@{}l}{\textbf{Compare to Existing Workflow}} \\
    Q25 & Pause to articulate rationales & 3.75 & 4.00 & 0.97 \\
    Q26 & Linking coding to positionality & 3.25 & 3.00 & 1.54 \\
    Q27 & Consider alternative interpretations & 3.67 & 4.00 & 1.15 \\
    Q28 & Re-read context before coding & 3.50 & 3.00 & 0.90 \\
    Q29 & Revisit earlier excerpts & 3.42 & 3.00 & 1.00 \\
    Q30 & Discuss disagreements productively & 3.75 & 4.00 & 0.87 \\
    Q31 & Revise code definition or split/merge codes & 4.00 & 4.00 & 0.74 \\
    \bottomrule
  \end{tabular}
\end{table*}

\newpage
\subsection{Interview Guide/Script}
\subsubsection{Formative Study}
\label{appendix:formative-interview-script}

``Thank you for participating in our study. Today, we'd like to discuss your experiences with qualitative data analysis. Specifically, we're interested in your coding practices, how you reflect on your interpretations, and how you collaborate with others.

This session will be recorded for analysis purposes, and all data will be anonymized. Please know that there are no right or wrong answers; we are genuinely interested in your personal experiences and thought processes.

Do you have any questions before we begin? Do we have your consent to start recording?''

\rule{\linewidth}{0.4pt} 

\noindent\textbf{II. Main Interview Questions}

\paragraph{Part 1: The ReflexiveLens}
\begin{enumerate}[label=\arabic*.]
    \item To start, could you recall your most recent or memorable qualitative data analysis project? Please briefly describe its setting, context, and the process you followed.
    
    \item Can you walk me through a time when you intentionally reflected on your own interpretation while you were coding?
    \begin{itemize}
        \item What triggered that reflection?
        \item Are there specific moments in the coding process where reflection feels especially necessary (e.g., when applying a new code, when you're unsure, when generating themes)?
    \end{itemize}
    
    \item Do you ask yourself certain kinds of questions to guide your interpretation of the data?
    \begin{itemize}
        \item If so, could you share an example?
        \item Do you typically document those thoughts, for instance, in a memo or a research journal?
    \end{itemize}
    
    \item In your own words, what specific actions or practices do you believe are essential for making a thematic analysis truly reflexive?
    
    \item Now, let's imagine a qualitative analysis tool designed to support reflection. Do you think it would be more helpful if such a tool provided reflective prompts \textit{proactively} (e.g., when it detects a pattern or a shift in your coding) or \textit{reactively} (e.g., only when you explicitly ask for help or seem stuck)?
    \begin{itemize}
        \item Can you imagine a situation where a proactive prompt would be particularly useful?
        \item What kind of signal or moment might make a proactive prompt feel helpful rather than intrusive? (e.g., applying a single code many times, switching between codes frequently, coding passages that seem very different with the same code).
        \item Conversely, are there situations where prompts would feel distracting?
        \item Ideally, who should be in control of when these prompts appear---you or the system?
    \end{itemize}
    
    \item \textit{[Show Concept 1: Reflective Prompts Interface]} I'd like to show you a concept from a system we're exploring. It’s designed to offer prompts during coding.
    \begin{itemize}
        \item Looking at these examples, what kinds of prompts would be most meaningful or helpful for you in your own work?
    \end{itemize}
\end{enumerate}

\paragraph{Part 2: Evolving Code Definitions}
\begin{enumerate}[label=\arabic*.]
    \item Can you recall a time when a code you created ended up meaning something different as the project went on?
    \begin{itemize}
        \item How did that shift occur? Was it gradual, or was there a specific moment of realization?
    \end{itemize}
    
    \item Have you ever realized a code was being overused or applied too broadly?
    \begin{itemize}
        \item What triggered that realization?
    \end{itemize}
    
    \item Could you describe situations that might lead you to \textit{split} a single code into multiple codes, or to \textit{merge} separate codes into one?
    
    \item How important is it for you to document these kinds of shifts in your codes? Why or why not?
    
    \item Would it be helpful for a tool to alert you when a code’s meaning or application seems to be drifting over time?
    
    \item \textit{[Show Concept 2: Code Evolution Tracker Interface]} Here’s a related concept---a system that tracks how a code’s usage evolves and visualizes its history (e.g., splits, merges, renames).
    \begin{itemize}
        \item How could you imagine using a feature like this? (e.g., for personal reflection, for team communication, for transparency in a final report?)
        \item In what ways might a feature like this \textit{not} work for your process?
    \end{itemize}
\end{enumerate}

\paragraph{Part 3: Positionality-Aware Collaboration}
\begin{enumerate}[label=\arabic*.]
    \item Can you share an example of a time when you and a collaborator applied different codes to the same passage of data?
    \begin{itemize}
        \item From your perspective, what was each person seeing in the data?
        \item How did you and your collaborator approach that difference in interpretation?
    \end{itemize}
    
    \item Thinking about your past projects, how has your own positionality (e.g., your background, profession, life experiences) shaped how you interpret data?
    \begin{itemize}
        \item Could you provide a specific example?
    \end{itemize}
    
    \item When trying to understand a collaborator’s interpretation, what kind of information is most helpful to you?
    
    \item In your collaborative projects, do you typically aim for absolute consistency in coding across team members? Why or why not?
    
    \item Do you feel that interpretive disagreements between collaborators should be \textit{resolved} to find a single consensus, or \textit{explored} as a source of richer insight?
    \begin{itemize}
        \item Can you share a moment where exploring a disagreement was especially productive?
        \item Can you share a moment where a disagreement was especially frustrating?
    \end{itemize}
\end{enumerate}

\noindent\textbf{III. Conclusion}
\begin{enumerate}[label=\arabic*.]
    \item Is there anything else we haven’t talked about that you think is important for understanding how qualitative researchers make sense of their data?
\end{enumerate}

\vspace{1em} 
``Thank you so much for your time and for sharing your valuable insights. This has been incredibly helpful for our research.''

\subsubsection{User Study Interview}
\label{appendix:user-study-interview-script}

\paragraph{0. Warm-up \& Initial Impressions (\textasciitilde45s)}
\begin{itemize}
    \item What were your immediate impressions when using the tool? What led to those impressions?
    \item How was the experience different from or similar to what you currently use or expected? 
    \item Did any part of the process feel particularly mentally demanding?
    \begin{itemize}
        \item \textit{Probe:} Was it a particular feature or a specific step in the workflow?
    \end{itemize}
\end{itemize}

\rule{\linewidth}{0.4pt}

\paragraph{1. ReflexiveLens Feature (\textasciitilde2 min)}
\textit{[Participants are shown a screenshot of the ReflexiveLens feature.]}
\begin{itemize}
    \item Were there any moments where you chose to use the reflection features \textit{before} applying a code to a passage? Could you tell us more about that?
    \begin{itemize}
        \item Why did you choose to reflect rather than immediately applying a code?
        \item How did these in-place reflections help you engage with the data or consider your positionality as a researcher?
        \item Did being able to reflect before coding help you manage cognitive load or the pressure of coding?
    \end{itemize}
    \item If you found this feature useful, have you done something similar in your past work? If so, how did you do it before?
\end{itemize}

\paragraph{2. Code History Feature (\textasciitilde2 min)}
\textit{[Participants are shown a screenshot of the Code History timeline.]}
\begin{itemize}
    \item Did you use, or could you see yourself using, the code development timelines?
    \begin{itemize}
        \item In what ways might this history help you as a researcher while you are developing codes?
        \item How could a feature like this help you communicate with your research team?
        \item Could this code history be useful when reporting your analysis results? Why?
    \end{itemize}
    \item If you found this feature useful, is tracking code history something you've done before? How?
\end{itemize}

\paragraph{3. Positionality-Aware Collaboration Feature (\textasciitilde2 min)}
\textit{[Participants are shown a screenshot of the disagreement resolution interface.]}
\begin{itemize}
    \item When you encountered a disagreement, how useful was it to see the reflection prompts and the positionality of your collaborator?
    \begin{itemize}
        \item Did this information change how you interpreted the data or their coding choice?
        \item How would you have approached this conversation with your collaborator?
    \end{itemize}
    \item If you found this feature useful, how have you navigated these kinds of interpretive differences in the past?
\end{itemize}

\paragraph{4. Living Concepts \& Code Drift Alerts (\textasciitilde2 min)}
\textit{[Participants are shown a screenshot of the Code Drift Alert.]}
\begin{itemize}
    \item Were there any moments when the code drift alert appeared? How did that make you feel?
    \begin{itemize}
        \item Did the alert seem to intervene at an accurate or helpful moment?
        \item What was your reaction when you saw the alert?
    \end{itemize}
    \item Can you describe a time the drift alert surfaced a potential shift in how you were using a code? How accurate was it, and what did you do next?
    \begin{itemize}
        \item Did this feature support the idea that codes are constructed and evolve, rather than simply being ``found''?
    \end{itemize}
    \item If you found this feature useful, have you ever tried to monitor for code drift before? How?
\end{itemize}

\paragraph{5. Discussion Focus Feature (\textasciitilde2 min)}
\textit{[Participants are shown a screenshot of the Discussion Focus Icon and Show Diff View Toggle.]}
\begin{itemize}
    \item How useful was it to see and set the discussion status for different disagreements? How did it guide your focus?
    \begin{itemize}
        \item How does this compare to your usual process for discussing coding differences with collaborators?
    \end{itemize}
    \item Compared to how you normally prioritize disagreements, what felt better or worse about this approach?
    \item Did the Discussion Focus feature facilitate more targeted discussions, or did it feel like a metric that contradicts the reflexive nature of the process?
    \item If you found this useful, how have you prioritized discussion topics with collaborators in the past?
\end{itemize}

\paragraph{6. Overall Workflow \& Concluding Thoughts (\textasciitilde2 min)}
\begin{itemize}
    \item Overall, could you see yourself using a system like this, or specific features from it, for your future qualitative analysis work? Why or why not?
\end{itemize}

\end{document}